\documentclass[journal]{IEEEtran}
\usepackage[labelfont=bf]{caption}

\usepackage{cite}
\usepackage{xcolor}

\usepackage{times}

\usepackage{soul}
\usepackage{url}
\usepackage[hidelinks]{hyperref}
\usepackage[utf8]{inputenc}
\usepackage{graphicx}
\usepackage{amsmath}
\usepackage{amsfonts}
\usepackage{booktabs}
\usepackage[flushleft]{threeparttable}
\usepackage{multirow}
\usepackage{tabularx}
\usepackage{xspace}
\usepackage{subcaption}
\usepackage{comment}
\usepackage{amsmath}
\usepackage{mathtools}
\usepackage{caption}
\usepackage{enumitem}
\usepackage{bbm}
\usepackage{wrapfig}
\usepackage{adjustbox}
\usepackage{varwidth}

\newcommand{\etal}{\mbox{\emph{et al.}}\xspace}

\newcommand{\baseline}{\mbox{$\mathop{\text{TptCF}}\limits$}\xspace}
\newcommand{\personlized}{\mbox{$\mathop{\text{PTN}}\limits$}\xspace}
\newcommand{\encounter}{\mbox{$\mathop{\text{CoPM}}\limits$}\xspace}
\newcommand{\logmean}{\mbox{$\mathop{\text{HCFM}}\limits$}\xspace}
\newcommand{\HAM}{\mbox{$\mathop{\text{HAM}}\limits$}\xspace}
\newcommand{\HGN}{\mbox{$\mathop{\text{HGN}}\limits$}\xspace}
\newcommand{\Caser}{\mbox{$\mathop{\text{Caser}}\limits$}\xspace}

\ifCLASSINFOpdf
\else
\fi

\begin{document}
%
\title{Hybrid Collaborative Filtering Models\\for Clinical Search Recommendation}

\author{Zhiyun Ren, 
        Bo Peng, 
        Titus K. Schleyer 
        and Xia Ning* 
\thanks{Zhiyun Ren is with the Department
of Biomedical Informatics, The Ohio State University, Columbus,
OH, 43210 USA e-mail: ren.685@osu.edu}
\thanks{Bo Peng is with the Department of Computer Science and Engineering, 
The Ohio State University, Columbus,
OH, 43210 USA e-mail: peng.707@buckeyemail.osu.edu.}
\thanks{Titus K. Schleyer is with Regenstrief Institute and Indiana University School of Medicine, 
Indianapolis, IN, 46202 USA e-mail: schleyer@regenstrief.org.}
\thanks{Xia Ning is with the Department
of Biomedical Informatics and the Department of Computer Science and Engineering, 
The Ohio State University, Columbus,
OH, 43210 USA e-mail: ning.104@osu.edu.} 
\thanks{*corresponding author}}

\markboth{Journal of \LaTeX\ Class Files,~Vol.~14, No.~8, August~2015}%
{Shell \MakeLowercase{\textit{et al.}}: Bare Demo of IEEEtran.cls for IEEE Journals}

\maketitle

\begin{abstract}

With increasing and extensive use of electronic health records, clinicians are often under time pressure when they need to retrieve 
important information efficiently among large amounts of patients' health records in clinics. 
While a search function can be a useful alternative to browsing through a patient's record, 
it is cumbersome for clinicians to search repeatedly for the same or similar information on similar patients. 
Under such circumstances,  there is a critical need to build effective recommender systems that can 
generate accurate search term recommendations for clinicians. 
In this manuscript, we developed a hybrid collaborative filtering model using patients'  
encounter and search term information 
to recommend the next search terms for clinicians to retrieve important information fast in clinics.  
%
%
%
%
For each patient, the model will recommend terms that either have high co-occurrence frequencies with his/her most recent ICD  codes 
or are highly relevant to the most recent search terms on this patient. 
We have conducted comprehensive experiments to evaluate the proposed model, and the experimental results 
demonstrate that our model can outperform all the state-of-the-art baseline methods for top-$N$ search term recommendation 
on different datasets. 
\end{abstract}


\IEEEpeerreviewmaketitle

\section{Introduction}
\label{sec:Introduction}
\IEEEPARstart{E}{lectronic} Health Records (EHRs) are increasingly large and varied collections of health information 
about patients. However, it is difficult for clinicians to retrieve information from EHRs efficiently and effectively given today's 
limitations of user interfaces. Clinicians often operate under time pressure, and must invest significant amounts of time and 
effort in retrieving information from EHRs. While using a search function can provide a useful alternative to browsing through 
a patient's record, searching repeatedly for the same or similar information on similar patients is repetitive and cumbersome. 
Under such circumstances,  there is a critical need to build effective systems that can generate accurate search 
term recommendations for clinicians. 

In this manuscript, we tackle the problem of next search term recommendation. 
Given each patient's previous encounters and previous search terms that clinicians conducted on the patient, the objective of the next search 
term recommendation problem is to recommend information items to clinicians that are most relevant to the patient to help diagnosis. 
We assume clinicians conduct search over the patient EHR data in order to manually identify such information 
items, while our method is able to proactively identify such information items and present them to the clinicians so as to
save the time and effort of manual search.

Recommender systems (RS) aim to recommend the most relevant items to a user  by 
learning user preferences from their previously interacted items (e.g., products, information items). 
Traditional RS techniques, such as collaborative filtering (CF)   
~\cite{ricci2011introduction, Ning2015}, 
have been widely used to recommend the top-$N$ most relevant items to each user -- the so-called top-$N$ recommendation. 
In this manuscript, we adapt traditional CF techniques and consider patients as users, 
search terms as items, and propose a RS model to recommend the next search term on each patient, and to present the 
terms to the clinicians. 
While traditional RS techniques usually recommend the next items only based on users' previously 
interacted items, the next search term recommendation can be a different problem as a search term on a patient  
may have  strong relations with two factors: previous search terms (e.g., for chronic diseases) 
and the patient's previous encounters (e.g., for most recent health concerns). 

Based on this intuition, we propose a model named 
\underline{H}ybrid \underline{C}ollaborative \underline{F}iltering \underline{M}odel using encounter information, denoted as \logmean, to recommend 
the next search term on a patient based on 
previous search terms on the patient and his/her previous encounters. 
Given the ICD codes in a patient's encounters and the previous search terms on each patient, 
in \logmean, we first calculate the co-occurrence frequency between each ICD code and each search term, where
a co-occurrence is considered when a term is searched right after the ICD  code has been 
assigned to the patient. 
The \logmean model then recommends terms that have high co-occurrence frequencies with the most recent ICD  codes  
and are highly relevant to the most recent search terms. 
%
In addition, we propose another model that 
uses only ICD-search term co-occurrence patterns to recommend the next search term for clinicians. 
This method  recommends the terms that are highly relevant to all the previous encounters of a patient, 
and is named 
\underline{C}o-\underline{o}ccurrence \underline{P}attern based recommendation \underline{M}ethod, denoted as \encounter.  
The experimental results show that the proposed models outperform all the state-of-the-art 
methods on top-$N$ search term recommendation on different real datasets.  
We have also conducted comprehensive parameter studies 
that enable important insights on the two factors. 
%

 

\section{Related Work}
\label{sec:related}
The most relevant work to our method is from recommender systems (RS) research field. 
Recently, a lot of research in RS has been conducted to tackle top-$N$ recommendation problems (i.e., to recommend
$N$ most relevant items). 
Specifically, factorized personalized Markov chains
(FPMC)~\cite{rendle2010factorizing} generate recommendations for the next item using Markov 
chains, which capture item-item transition relations. 
%
%
Hidasi \etal~\cite{hidasi2015session,hidasi2018recurrent} adapted deep learning 
techniques, and used gated recurrent  units (GRUs) to model the dynamics of a  
user's preferences.  
Kang \etal~\cite{kang2018self} developed a self-attention based sequential 
model (SASRec) to capture the  most informative items in a user's previously 
interacted items for top-$N$ recommendation.  
Tang \etal~{\cite{tang2018personalized}} developed 
a convolutional sequence embedding recommendation model (\Caser) 
by adapting multiple convolutional filters on the most recent interacted items of each user 
to model sequential features in a user's historical interactions.   
Ma \etal~\cite{ma2019hierarchical} developed a hierarchical gating network (\HGN) 
that adapts gating mechanisms over users' historical interacted items to identify 
important items and their features  for top-$N$ recommendation.

Very limited work has been conducted on next search term recommendation problem for clinical applications. 
Fan \etal~\cite{fan2019improving} have developed several CF  
models to tackle this problem. 
Among these models, Transition-Involved Patient-Term-Similarity-based CF Scoring (\baseline) 
recommends the next search terms for clinicians on a patient based on the patient-patient similarities, 
term-term similarities and term-term transition relations. 
The difference between \baseline and our model is that our model generates top-$N$ recommendations 
using both previous search terms and previous encounters, while \baseline only uses  
previous search terms. 

\section{Definitions and Preprocessing}
\label{sec:notations}

\subsection{Definitions and Notations}
\label{subsec:definitions}
\begin{figure}[!t]
\vspace{-5pt}
\centering
    \includegraphics[width=1\linewidth]{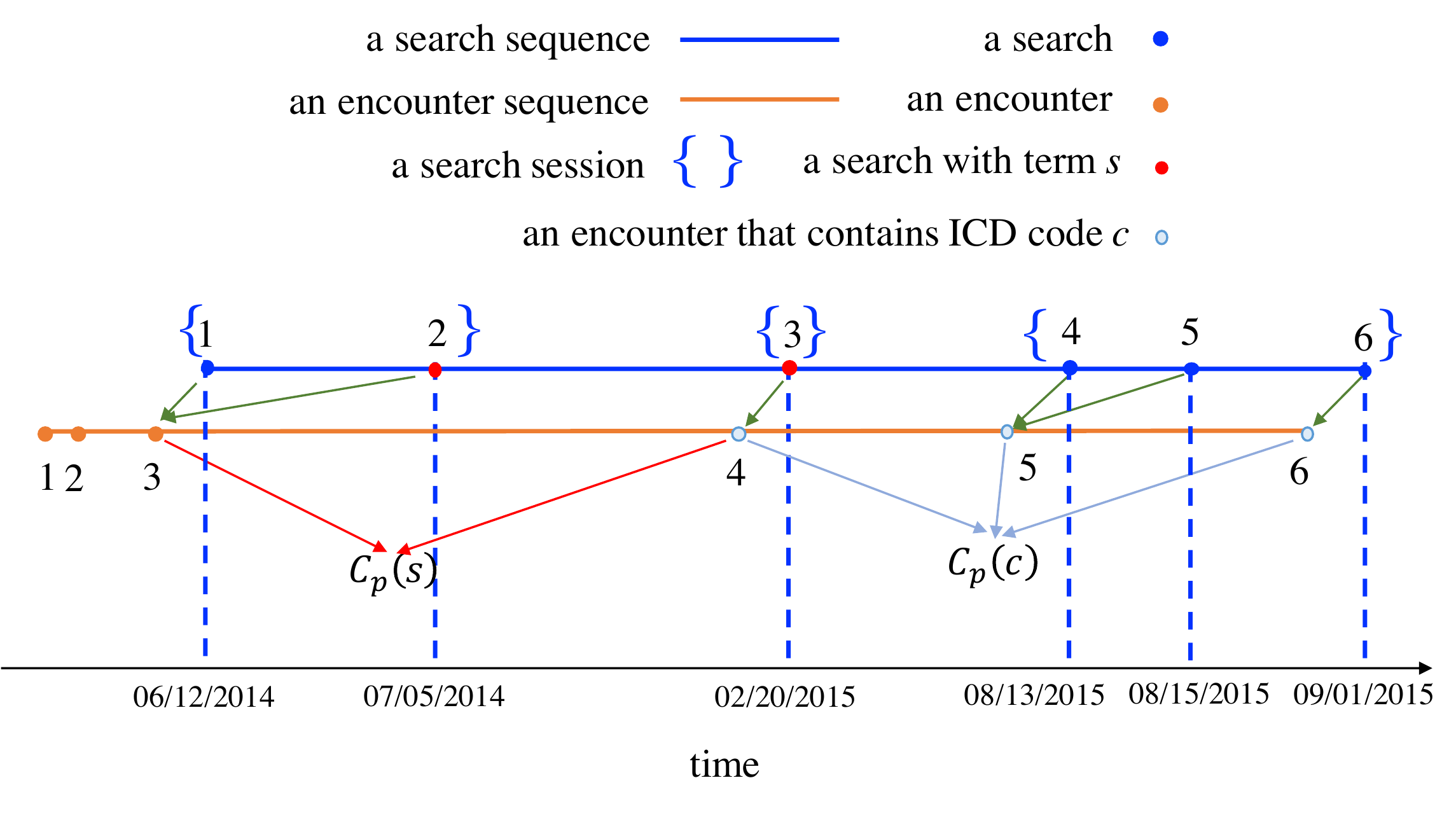}
    \caption{{Data Preprocessing Protocols}}
    \label{fig:matching}
    \vspace{-5pt}
\end{figure}

Figure \ref{fig:matching} presents the  data preprocessing protocols and 
Table~\ref{tbl:notations} presents the key notations used in this manuscript.  
Formally, 
the terms searched on each patient will be sorted chronologically. 
The sequence of patient $p$'s sorted search terms is denoted as $S_p$, and 
the subsequence of $S_p$ from 
the $i$-th search to the $j$-th search is denoted as $S_p(i, j)$. 
For simplicity, we store in $S_p$ the indices of search terms from a universal dictionary instead of search terms themselves. 
Similar to search terms, the encounters of each patient will be sorted chronologically.  
The sequence of patient $p$'s sorted encounters is denoted as $C_p$, and  
the subsequence of $C_p$ from
the $i$-th encounter to the $j$-th encounter are denoted as $C_p(i, j)$.  
For each patient, each search will be matched using timestamps to his/her most recent, previous encounter that the search 
immediately follows (indicated by the green arrows in Figure~\ref{fig:matching}). 
Please note that such matching does not necessarily mean that the searches happen during 
the matched encounters, 
or they are triggered or induced by the encounters, but it only indicates the temporal proximity. 

%

For each patient, an ICD code may appear in his/her multiple encounters. 
We denote the encounters of patient $p$ that contain ICD code $c$ as $C_p(c)$.  
Similarly, a term can be searched for multiple times on a patient. 
We denote the encounters of patient $p$ that each search term $s$ is matched to as $C_p(s)$.  
%
We use the indices of ICD codes or search terms to refer them correspondingly in the sequences. 

\subsection{Identifying Sessions}

In our dataset (discussed later in Section~\ref{sec:materials:dataset}), the time difference 
between consecutive searches in a search sequence may vary from minutes to years, but 
unfortunately session information is not always logged.  
Therefore, we segment searches into sessions based on their timestamps using a sliding window 
with size of three months. Therefore, 
if the time interval between two consecutive searches are less than three months, the two search terms will be 
grouped into one session. 
To generate recommendations for a search, 
we  only use the searches that are in a same session at the time when the recommendation 
is to be made. 
This is analogous to the real patient visit scenario in which clinicians search within the context of the 
 current visit or recent searches. 
Figure \ref{fig:matching} shows the search sessions with blue parentheses. 
%
Please note that the sessions here are defined as a natural segmentation of search/encounter clusters, and they are 
not the same as the sessions in internet connections.

\begin{table}[!t]
  \caption{Notations}
  \label{tbl:notations}
  \centering
  \vspace{-5pt}
  \begin{threeparttable}
      \begin{tabular}{
	@{\hspace{0pt}}l@{\hspace{2pt}}
	@{\hspace{2pt}}p{0.85\linewidth}@{\hspace{0pt}}          
	}
        \toprule
        notations & meanings \\
        \midrule
        $S_p$ & the sequence of patient $p$'s chronologically sorted searches \\
        $S_p(i, j )$ & subsequence of $S_p$ from the $i$-th search to the $j$-th search  \\
        $C_p$ & the sequence of patient $p$'s chronologically sorted encounters \\
        $C_p(i, j )$ & subsequence of $C_p$ from the $i$-th encounter to the $j$-th encounter  \\
        $C_p(c)$ & the encounters of patient $p$ that contain ICD code $c$  \\
        $C_p(s)$ & the encounters of patient $p$ that search term $s$ is matched to  \vspace{0.5em}\\
	
        $n$/$m$/$l$ &  number of ICD codes/search terms/patients \\
        $d$ &  the dimension of representations \\
        $n_p$/$l_p$  & the number of all search terms/encounters on patient $p$ at the time when the  
                     recommendation is to be made \\
         
        $m_s$/$m_c$ & the number of previous search terms/encounters that are used for next 
                    search term recommendation \\
       \bottomrule
      \end{tabular}
  \end{threeparttable}
\end{table}


\section{Methods}
\label{sec:method}

Our method has the following steps. 
First, we learn to numerically represent each ICD code and each search term so that the representations 
can be used to calculate recommendation scores. Such representations will encode how ICD codes and 
search terms co-occur in our data (Section~\ref{sec:method:learning}).  
Then, we calculate a recommendation score for each term for each patient. 
The recommendation score will be based on the previous search terms on the patients, 
and the previous encounters of the patients. 
We rank the recommendation scores of all terms for each patient and recommend top-$N$ terms 
to the clinician seeing the patient (Section~\ref{sec:method:recommendation:EiCF}).

\subsection{Learning from Previous Search Terms and Encounters}
\label{sec:method:learning}

\subsubsection{Constructing an ICD Code-Search Term Co-Occurrence Matrix}
\label{sec:method:learning:construct}
%
We hypothesize that each search term on a patient is highly related to the patient's most recent 
encounters, that is, given an ICD  code that is assigned to a patient, and thus the corresponding diagnoses, 
certain terms that are related to the encounters/ICD codes are more likely to be searched next.   
For example, if a patient was assigned an ICD code ``588.81: secondary hyperparathyroidism 
(of renal origin)" in his/her recent encounters, terms such as potassium levels that are highly related to hyperparathyroidism 
have high probability to follow compared to the case of ICD code ``786.2: Cough"
with which potassium level provides little information. 
Thus, co-occurrence frequencies between ICD codes and search terms can provide  
useful information to predict search terms that will follow;  
given the ICD  codes in a patient's recent encounters, the terms with high co-occurrence 
frequencies with these ICD  codes among many patients are more likely to be searched next, and thus should be 
recommended. 

Following the above intuition, we first calculate the co-occurrence frequency between  
each ICD  code and each search term
by counting how many times in total the term has been searched after the ICD code in encounters. 
We construct a matrix $A \in \mathbb{R}^{n \times m}$ to store such co-occurrence frequencies,  
where $n$ is the number of all involved ICD  codes and $m$ is the number of all search terms.   
We hypothesize that clinicians tend to search information from recent encounters of a  
patient. Thus, 
a search term is more likely to be induced from its recent ICD codes compared to 
those from long time ago. 
%
Based on this, we emphasize the recent encounters using 
a time-decay parameter, and calculate the ICD code-search term co-occurrence frequencies $a_{cs}$ as follows:  
\begin{equation}
    \label{eqn:cooccur}
    \begin{aligned}
    a_{cs} = \sum_{p=1}^{l}  \   \sum_{{e_c} \in C_p(c)} \   \sum_{{e_s} \in C_p(s)}  \lambda^{i(e_s)-i(e_c)} \mathbbm{1} (i(e_s) \geq i(e_c)) ,
    \end{aligned}
\end{equation}
where $e_s$ and $e_c$ are two encounters; 
$l$ is the total number of patients, 
$\lambda \in (0,1)$ is the time-decay parameter (in our experiments, $\lambda=0.5$); 
$\mathbbm{1}(x)$ is the indicator function ($\mathbbm{1}(x) = 1$ if $x$ is true, otherwise, $\mathbbm{1}(x) = 0$); 
$i(e_s)$ and $i(e_c)$ are the indices of encounter $e_s$ and encounter $e_c$, respectively, in patient $p$'s encounter
sequence $C_p$. 
When calculating the co-occurrence frequencies between ICD code $c$ and term $s$, we only consider the 
cases in which  term $s$ is searched after ICD code $c$ or in the same encounter as $c$ 
(i.e., $\mathbbm{1} (i(e_s) \geq i(e_c))$).
%
%
%
Please note that $a_{cs}$ is not a probability value and can have values greater than 1; larger $a_{cs}$ indicates 
more likely that ICD code $c$ and search term $s$ co-occur. 
%

\subsubsection{Learning ICD Code and Search Term Representations}
\label{sec:method:learning:latent}
Note that the co-occurrence matrix $A$ as constructed above is typically sparse because most of the ICD codes are not 
observed as co-occurring with most of the search terms. 
In order to capture the underlying relations between each ICD code and each search term that are not directly observed 
in $A$, 
we use a matrix factorization method~\cite{koren2009matrix} to learn the representations of ICD codes and search terms  
which together produce what we observe in matrix $A$ and recover what we do not observe in $A$.  
Specifically, we factorize $A$ into two low-rank matrices, 
$U \in \mathbb{R}^{n \times d}$ and $V \in \mathbb{R}^{m \times d}$ ($d < \min(n, m)$), 
representing the ICD  codes and search terms, respectively. 
Particularly, each row of matrix $U$, denoted as $\mathbf{u}_c$, represents 
the ICD  code $c$, 
and each row of matrix $V$, denoted as $\mathbf{v}_s$, represents  
the search term $s$. 
Thus, all ICD  codes and search terms are represented by size-$d$ latent vectors and can be learned 
from matrix $A$. 
%
The co-occurrence ``chance" between ICD  code $c$ and search term $s$ can be estimated as follows:
\begin{equation}
    \label{eqn:factorize_cooccur}
    \hat{a}_{cs} = \mathbf{u}_c \mathbf{v}_s^{\mathsf{T}}, 
\end{equation}
where $\hat{a}_{cs}$ is the estimation of ${a}_{cs}$. 
To learn the representations of each ICD code and each search term, 
we formulate the following optimization problem:  
\begin{equation}
    \label{eqn:factorize_opt}
    \underset{U, V}{\min}\  \|A - UV^{\mathsf{T}}\|_F^2 + \frac{\gamma}{2}(\|U\|_F^2 + \|V\|_F^2), 
\end{equation}
where $U = [\mathbf{u}_1; \mathbf{u}_2; \cdots; \mathbf{u}_n]$, 
$V = [\mathbf{v}_1; \mathbf{v}_2; \cdots; \mathbf{v}_m]$, 
$\gamma$ is the weight for regularization term; 
$\|.\|_F$ is the Frobenius norm, and the regularization on the Frobenius norm restricts large values in $U$ and $V$. 
We  solve this problem using alternative gradient descent method \cite{koren2009matrix}. 
Other representation learning methods include deep learning based methods~\cite{tang2018personalized,kang2018self}. 
However, given the very sparse nature of our data (Section~\ref{sec:materials:dataset}), we do not use the deep learning 
methods which typically require a lot of data for model training. Instead, matrix factorization based method as in 
Equation~\ref{eqn:factorize_cooccur} has been demonstrated very effective in learning from sparse data~\cite{koren2009matrix}.


\subsection{\logmean: Hybrid Collaborative Filtering Model}
\label{sec:method:recommendation:EiCF}


Our hybrid collaborative filtering model using encounter information, 
denoted as \logmean, generates recommendations for the next search term for clinicians on each patient 
using two factors: 
(i) the previous search terms on the patient, and 
(ii) the previous encounters of the patient. 
Each of two factors will be used to calculate a recommendation score for each term for the next 
search on a patient (Section~\ref{sec:method:recommendation:EiCF:prevTerm} and Section~\ref{sec:method:recommendation:EiCF:prevEnc}), 
and the two recommendation scores will be combined into a final recommendation score (Section~\ref{sec:method:recommendation:EiCF:rec}). 


\subsubsection{{Recommendation Score from Previous Search Terms}}
\label{sec:method:recommendation:EiCF:prevTerm}

Previous studies from RS field have shown that more recent items will provide more 
pertinent information to recommend the next item \cite{tang2018personalized,ma2019hierarchical}. 
We hypothesize similarly that most recent search terms provide more pertinent information to recommend the
next search term. 
Based on this, we recommend the next term using the 
most recent search terms on a patient. 
%
%
We aggregate the information from the most recent $m_s$ search terms in the current search session
by calculating the mean values of their 
latent feature representations as follows:  
\begin{equation}
\begin{aligned}
    \label{eqn:meanprevterm}
    \mathbf{m}_p = \frac{1}{m_s}\sum_{i \in S_p(n_p-m_s,n_p)} \mathbf{v}_i,
\end{aligned}
\end{equation}
%
where 
$n_p$ is the number of all  search terms on patient $p$ at the time when the recommendation is to be made;  
$m_s$ is the number of the most recent search terms that are used for recommendation  
($m_s$ is a fixed number in our experiments);  
$\mathbf{m}_p  \in \mathbb{R}^{1 \times d}$ is the aggregated representation from previous $m_s$ 
search terms on patient $p$;   
%
%
%
The recommendation score of term $s$ on patient $p$ based on previous search terms 
is calculated as the dot-product similarity between $\mathbf{m}_p$ and $\mathbf{v}_s$ as follows: 
\begin{equation}
    \label{eqn:prevterm}
    {x}_{ps} = \mathbf{m}_p \mathbf{v}_s^{\mathsf{T}}. 
\end{equation}
%

\subsubsection{{Recommendation Score from Previous Encounters}}
\label{sec:method:recommendation:EiCF:prevEnc}
To calculate the recommendation score for each term, we also use  the information from the most recent 
$m_c$ encounters of each patient.  
%
%
%
Here, we hypothesize that clinicians tend to search within the context of the recent searches; 
ICD codes that induce the most recent searches  are more likely to induce the next search, and thus 
should be emphasized to recommend the next search term. 
Based on this, 
we learn an importance weight on each ICD code $c$ for each patient $p$. 
%
The importance weight is calculated as
the normalized dot-product similarity between each ICD  code and the most recent $m_s$ search terms 
as follows: 
\begin{equation}
\begin{aligned}
    \label{eqn:wjs}
     w_{pc}  
             =  \frac{\exp(\mathbf{u}_c \mathbf{m}_p^{\mathsf{T}} )}{{\sum_{e' \in C_p(l_p-m_c,l_p)}  \  \sum_{c' \in e'} \exp(\mathbf{u}_c' \mathbf{m}_p^{\mathsf{T}} )}}
\end{aligned}
\end{equation}
%
where $\mathbf{m}_p$ is calculated as in Equation~\ref{eqn:meanprevterm}; 
$l_p$ is the number of all encounters of patient $p$ at the time when the recommendation is to be made; 
$m_c$ is the number of the most recent encounters that are used for recommendation 
($m_c$ is a fixed number in our experiments); 
$e'$ is an encounter in $C_p(l_p-m_c, l_p)$, and $c'$ is an ICD code in $e'$. 
%
%
%
The recommendation score of term $s$ on patient $p$ based on previous encounters is calculated as follows: 
\begin{equation}
    \label{eqn:prevenc}
    {y}_{ps} = \sum_{e \in C_p(l_p-m_c, l_p)} \  \sum_{c \in e}  w_{pc} \mathbf{u}_c \  \mathbf{v}_s^{\mathsf{T}}, 
\end{equation}
%
%
where $e$ is an encounter in $C_p(l_p-m_c, l_p)$, and $c$ is an ICD code in $e$. 

\subsubsection{{Combination of Recommendation Scores} }
\label{sec:method:recommendation:EiCF:rec}
The recommendation scores of term $s$ on patient $p$ calculated as above are then combined into a final recommendation 
score as follows: 
%
\begin{equation}
    \label{eqn:recommendation}
    {r}_{ps} = \alpha \  {x}_{ps} + (1-\alpha) \  {y}_{ps} 
\end{equation}
where $\alpha \in [0,1]$ is a pre-defined weight for the two factors. 
In Equation~\ref{eqn:recommendation}, $\alpha$=$1$ indicates   
that only previous search terms are used for recommendation, 
and $\alpha$=$0$ indicates that only  previous encounters are used for recommendation.  
The recommendation scores of all  terms will be sorted, and the  terms with  
top-$N$ scores will be recommended for the next search. 

\subsection{\encounter: Co-occurrence Pattern-based Model}
\label{sec:method:recommendation:CMF}
%

We propose another method, denoted as \encounter, that 
uses only ICD-search term co-occurrence patterns (Section~\ref{sec:method:learning}) 
to recommend the next search term for clinicians. 
The difference between \logmean and \encounter is that 
\logmean recommends the search terms that are most relevant to the most recent 
search terms and the most recent encounters, 
whereas \encounter recommends the search terms that are most relevant to 
all the previous encounters of a patient. 
Both methods represent ICD codes and search terms using the representation matrices, $U$ and $V$, respectively, 
as learned based on Equation \ref{eqn:factorize_opt}. 
%
The \encounter model will serve as a baseline method in our experiments.

Specifically, we aggregate the information of all ICD codes  
in the patient's previous encounters  to calculate the recommendation score for each search term. 
We hypothesize that more recent ICD codes are more likely to induce the next search term. Therefore, 
we emphasize the encounters/ICD codes that are closer to a search to generate the term 
recommendations for this search  using a time-decay parameter. 
%
%
%
Specifically, 
the recommendation score for term $s$ on  patient $p$ is calculated as follows: 
\begin{equation}
    \label{eqn:factorize_cooccur_pred}
     r_{ps} =  \sum_{e \in C_p(1,l_p)} \sum_{c \in e} \sigma^{i(e_s)-i(e)} \mathbf{u}_c \mathbf{v}_s^{\mathsf{T}}, 
\end{equation}
%
where $e$ is an encounter in $C_p(1, l_p)$, and $c$ is an ICD code in $e$; 
$e_s$ is the encounter that current search is matched to, that is, the most recent, previous encounter 
at the time that the recommendation is to be made;  
$i(e_s)$ and $i(e)$ are the indices of encounter $e_s$ and encounter $e$, respectively; 
$\sigma \in (0,1)$ is the time-decay parameter (in our experiments, $\sigma$=$0.5$). 
%
%
Note that the time-decay parameter $\sigma$ here indicates how long ago each encounter occurred before the 
time of recommendation, whereas the time-decay weights $\lambda$ in Equation~\ref{eqn:cooccur} indicates  
the temporal proximity between an encounter and a search term. 
Thus, the two time-decay parameters represent different information in the model. 
Finally, we  sort the recommendation scores of all the terms and recommend the 
top-$N$ terms for the next search.

\section{Materials}
\label{sec:materials}

\subsection{Baseline Methods}
\label{sec:materials:baseline}
We compare the \logmean and \encounter with the following state-of-the-art baseline methods. 
Some of the methods are from recommender systems (RS) research, in which 
they have users and items. 
To use these methods on the next search term recommendation task, 
we consider patients analogous to users, and search terms analogous to items in our 
experiments.

\begin{itemize}[noitemsep,nolistsep,leftmargin=*]
\item \textbf{Transition-Involved Patient-Term-Similarity-based CF Scoring (\baseline)} 
\cite{fan2019improving}. 
The \baseline model generates recommendations for the next search term using two factors: 
(1) patients' similarities and search terms' similarities (similarity-based scoring) and 
(2) search term dynamic transitions (dynamics-based scoring). 
For each patient, \baseline will calculate a recommendation score for each search term using 
similarity-based scoring and  dynamics-based scoring, and recommend the top search terms 
with the highest scores. 
The detailed explanation of the \baseline method will be presented in the supporting document. 

%
%
%
\item \textbf{Personalized top-$N$ (\personlized)} 
The \personlized model 
recommends the next item using the most frequently interacted items in each user's history. 
In our experiments,  
the terms are ranked based on how many times they are searched on each patient, and the patient's
top-$N$ most frequently searched terms will be recommended to the patient. 
%
%
%
\item \textbf{Hierarchical Gating Network (\HGN)}~\cite{ma2019hierarchical}. 
\HGN selects important items and their important features by adapting gating 
mechanisms \cite{ma2019hierarchical} over users' historical interacted items. 
To generate recommendations for the next items, 
\HGN uses the identified important items  and important features, and 
users' general preferences to calculate a recommendation score of each item for a user. 
\HGN is a state-of-the-art sequential recommendation method. 
%
\item \textbf{Hybrid Associations Model (\HAM)}~\cite{Peng2020HAM}. 
\HAM models users' long-term preferences from all their historical items, 
and models users' short-term preferences from their most recent items.  
The short-term preferences contain both high-order and low-order association patterns among items.  
Both the long-term and short-term preferences are used to recommend the next 
items. 
\HAM has been demonstrated as the state-of-the-art algorithm for sequential top-$N$ recommendation. 
%
%

\end{itemize}




\subsection{Datasets}
\label{sec:materials:dataset}

%
The data used in our experiments are collected from Eskenazi Health organization in Indiana, US, 
from 04/2013 to 05/2016.  
This dataset contains 13,934 patients, their 1,377,381 encounters, 9,565 valid ICD 9 codes 
and 7,215 unique search terms. 
%
Among the 7,215 unique search terms, we remove the irregular search terms, 
such as numbers and punctuations, and also infrequent search terms that appear only once, 
and map the misspelled terms to their most similar  
terms (e.g., ``adimssion" is mapped to ``admission"). 
We also remove the infrequent patients and keep those who have at least two search terms 
and at least three encounters. 
Table \ref{tbl:dataset} presents the statistics of the dataset after data preprocessing. 
Overall, 2,955 patients and 2,101 unique search terms are retained in the 
dataset. 
On average, each 
patient has 10.22  searches 
and 173.26 encounters. 
Each term has been searched for 14.37 times on average over all the patients, and each encounter has 2.09 ICD  codes on average.  
Our study is conducted under Protocol 
\#1612682149 ``Supporting information retrieval in the ED through collaborative filtering"
approved by the Indiana University Institutional Review Board (IRB). 

\begin{table}[h]
\small
  \caption{Dataset Statistics}
  \centering
  \vspace{-3pt}
  \label{tbl:dataset}
  \begin{threeparttable}
      \begin{tabular}{
        @{\hspace{0pt}}l@{\hspace{2pt}}
        @{\hspace{2pt}}r@{\hspace{2pt}}
        }
        \toprule
        Variables & Statistics \\
        \midrule
        Number of patients & 2,955 \\
        Number of {unique} search terms & 2,101 \\
        Number of {unique} ICD 9 codes & 7,027 \\
        Number of encounters & 511,987 \\
        Number of sessions & 3,488 \\
        \midrule
        Average number of {searches} per patient & 10.22 \\
        Average number of unique search terms per patient & 7.00 \\
        Average number of encounters per patient & 173.26 \\
        Average number of sessions per patient & 1.18 \\
        Average number of searches per term & 14.37 \\
        Average number of previous encounters per term & 228.89 \\
        Average number of search records per session & 8.66 \\
        Average number of unique ICD 9 codes per encounter & 2.09 \\
        \bottomrule
      \end{tabular}

  \end{threeparttable}
  \end{table}

\begin{figure}[h!]
\centering
    \centering
    \includegraphics[width=0.75\linewidth]{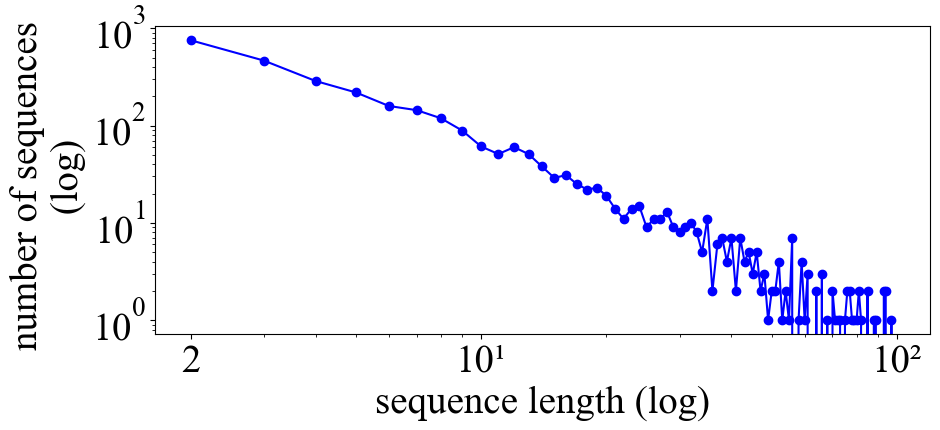}
    \caption{Distribution of Search Sequence Lengths}
    \label{fig:seq_len}
\end{figure}

\begin{figure}[!h]
    \centering
    \includegraphics[width=0.75\linewidth]{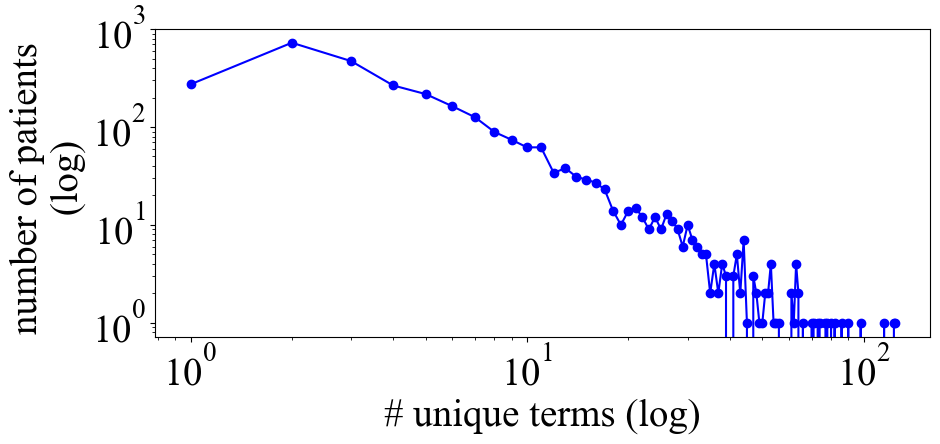}
   \caption{Distribution of Unique Terms per Patient}
    \label{fig:term_freq}
\end{figure}

We present the distribution of search sequences' lengths in the dataset in Figure~\ref{fig:seq_len}. 
%
As Figure~\ref{fig:seq_len} shows, there are many more short sequences in 
the dataset  than long sequences. 
Figure~\ref{fig:term_freq} presents the distribution of the number of unique terms for each patient. As it indicates, 
most of the search terms are not frequent. 
On average, each patient has 7.00 unique search terms. 
Moreover, as presented in Section \ref{subsec:definitions}, we group search terms into 
 sessions. 
Overall, there are 3,488 sessions identified from the dataset. 
On average, each patient has 1.18 sessions, and each session has 8.66 search terms. 
It is notable that search sequences are typically very short, and the number of unique search terms 
per patient is very small. This will make the recommendation problem difficult because of the 
sparsity of the available data. 
Table \ref{tbl:freq-terms} lists the most frequently searched terms, 
where frequency is calculated based on how many times each term is searched 
on all patients. 
%

\begin{table}[!h]
\centering
\footnotesize
  \caption{Most Frequent Search Terms}
  \centering
  \label{tbl:freq-terms}
  \begin{threeparttable}
      \begin{tabular}{
	@{\hspace{3pt}}r@{\hspace{3pt}}
	@{\hspace{3pt}}r@{\hspace{3pt}} |
	@{\hspace{3pt}}r@{\hspace{3pt}}
	@{\hspace{3pt}}r@{\hspace{3pt}} |
	@{\hspace{3pt}}r@{\hspace{3pt}}
	@{\hspace{3pt}}r@{\hspace{3pt}}
	}
        \toprule
         Term & Frequency   &   Term & Frequency   &   Term & Frequency \\ 

        \midrule
    
hiv & 2,144                    &            culture & 456 		&	urine & 284 \\
a1c & 854              	    &            inr & 431 		&	pap & 279 \\
creatinine & 709 		&	colonoscopy & 409 	&	height & 270 \\
weight & 706 			&	tsh & 401 			&	hgb & 253 \\
ekg & 634 			&	troponin & 371 		&	ct & 224 \\
cbc & 542 			&	ldl & 352 			&	urology & 221 \\
bmp & 475 			&	echo & 316 		\\



        \bottomrule
      \end{tabular}
       \end{threeparttable}
\end{table}

\subsection{Experimental Protocols}
\label{sec:materials:setting}
\begin{figure}[!h]
    \includegraphics[width=\linewidth]{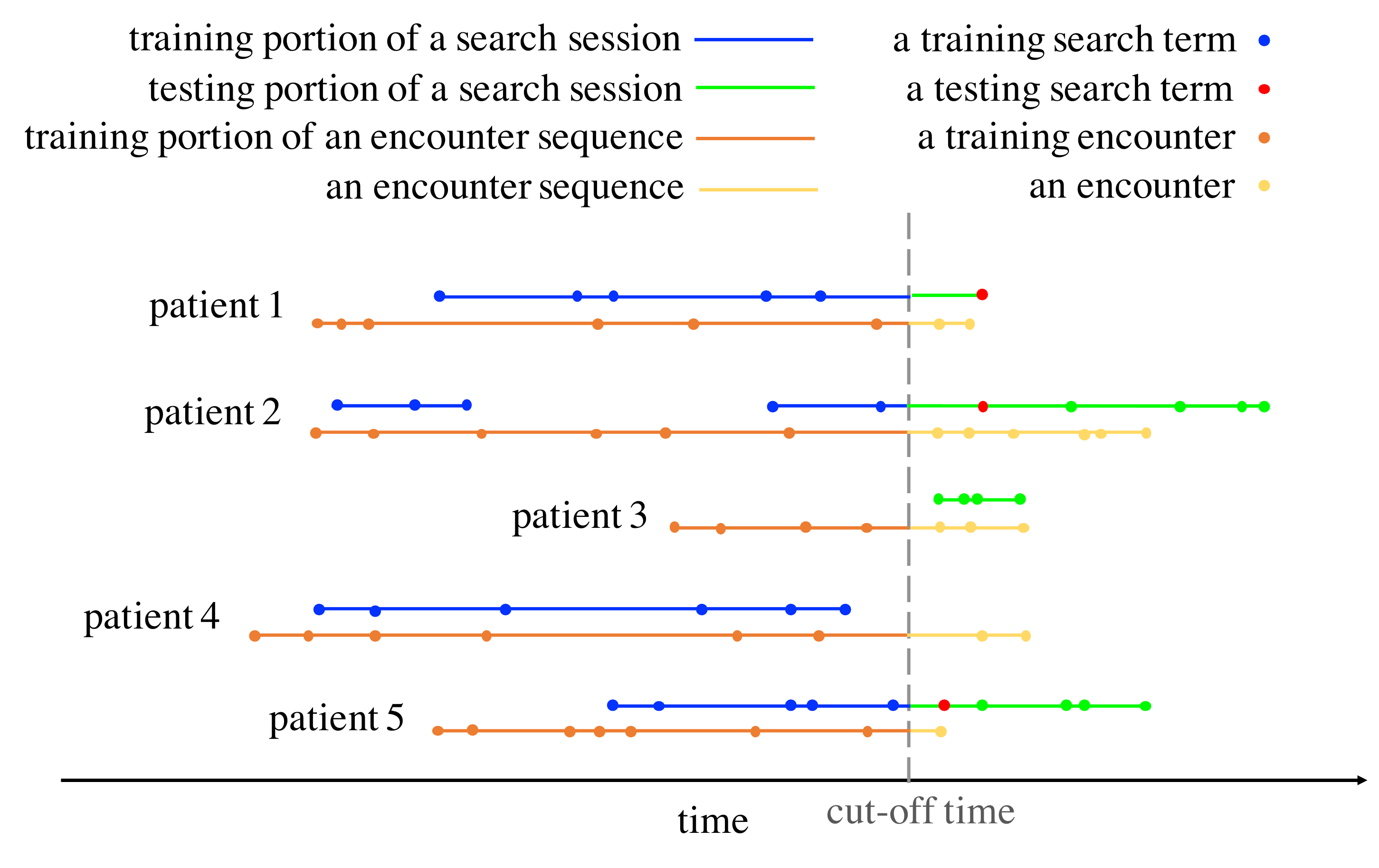}
    \caption{CUTOFF Experimental Protocol}
    \label{fig:setting}
\end{figure}

\begin{table}[h]
\small
  \caption{Dataset Statistics on Different Cut-off Times}
  \centering
  \label{tbl:dataset_cutoff}
  \begin{threeparttable}
      \begin{footnotesize}
      \begin{tabular}{
        @{\hspace{3pt}}l@{\hspace{6pt}}
        @{\hspace{6pt}}r@{\hspace{6pt}}
        @{\hspace{6pt}}r@{\hspace{6pt}}
        @{\hspace{6pt}}r@{\hspace{6pt}}
        @{\hspace{6pt}}r@{\hspace{6pt}}
        @{\hspace{6pt}}r@{\hspace{6pt}}
        @{\hspace{6pt}}r@{\hspace{6pt}}
        @{\hspace{6pt}}r@{\hspace{3pt}}
        }
        \toprule
        Cut-off Time & P$_{t}$ & P$_{e}$ & T$_{t}$ & T$_{e}$ & S$_{t}$ &T$_{t}$/S$_{t}$  & E$_{t}$/P$_{t}$  \\
        \midrule
        04/01/2015 &  2,627 & 109 & 1,915 & 442 & 2,854 & 8.22 & 167.88 \\ 
        10/01/2014 &  2,086 & 20  & 1,493 & 56 & 2,224 & 6.30 & 167.37 \\ 
        04/01/2014 &  1,129 & 139 & 1,065 & 254 & 1,171 & 6.12 & 164.84 \\ 
        10/01/2013 &  248   & 11  & 373   & 32 & 249 & 4.57 & 178.84  \\ 

        \bottomrule
      \end{tabular}
      \end{footnotesize}

      \begin{tablenotes}[normal,flushleft]
        \begin{footnotesize}
        \item
	      In this table, P$_{t}$ indicates the number of patients in the training set; 
              P$_{e}$ is the number of patients in the testing set; 
	      T$_{t}$ is the number of unique search terms in the training set; 
              T$_{e}$ is the number of unique search terms in the testing set; 
            
              S$_{t}$ is the number of search term sessions in the training set; 
              T$_{t}$/S$_{t}$ is the average number of search terms per session in the training set; 
              E$_{t}$/P$_{t}$ is the average number of encounters per patient in the training set. 
               
        \par
        \end{footnotesize}
      \end{tablenotes}
  \end{threeparttable}
\end{table}


We use the following experimental protocol to evaluate our methods on the dataset. 
In this protocol, all the search sequences are split by a same cut-off time. 
Any searches before the cut-off time are split into training set, and any searches after the cut-off time 
are split into testing set. The models are trained using only the training set, for example, the co-occurrence 
matrix (Equation \ref{eqn:cooccur}) is constructed only using the search terms and encounters in the training set. 
This protocol is referred to as cut-off cross validation, denoted as CUTOFF. 
Figure \ref{fig:setting} demonstrates the CUTOFF experimental protocol.
%

We selected four cut-off times: 10/01/2013, 04/01/2014, 10/01/2014 and 04/01/2015. 
These cut-off times are selected because sufficient search terms and encounters from a majority of the search sequences 
are retained in training set before the cut-off time and meanwhile sufficient search sequences have testing terms after the
cut-off time. After the split, the statistics for the training and testing data of four cut-off times that we choose 
are presented in Table \ref{tbl:dataset_cutoff}. This CUTOFF setting is close 
to the realistic scenario, that is, all the data before a certain time should be used to predict information after that time. 
However, a shortcoming of CUTOFF is that many early search sequences may not have testing terms, and many late 
search sequences will not have anything in training set. Sequences that do not have testing terms are still used 
to train models. Sequences that do not have training terms are not used. For those sequences which have terms 
after the cut-off time, only the first one of the terms after the cut-off time will be used for evaluation.

%

\subsection{Evaluation Metrics}
\label{sec:materials:metric}
We use hit rate at $k$ (HR@$k$) to evaluate the different methods. 
HR@$k$ measures the ratio of the patients whose ground-truth next search term is within the 
top-$k$ recommendations, over all patients. 
%
Higher HR@$k$ indicates better recommendation performance.

\subsection{Parameter Tuning}
\label{sec:materials:parameter}
%
We use grid search to tune four parameters: 
$m_s$ (Equation \ref{eqn:meanprevterm}), $m_c$ (Equation \ref{eqn:wjs}, \ref{eqn:prevenc}), 
 $\alpha$ (Equation \ref{eqn:recommendation}) and latent dimension $d$. 
%
%
We will present the best results among different parameter settings in Section \ref{sec:results}.

\section{Results and Discussions}
\label{sec:results}



\subsection{Overall Performance}
\label{sec:results:overall}

We compare \HGN, \HAM,  \baseline, \personlized, \encounter and \logmean in our experiments. 
Table \ref{tbl:results20150401} , \ref{tbl:results20141001}, \ref{tbl:results20140401} and \ref{tbl:results20131001} 
present the best performance of each method in terms of HR@$k$ ($k \in \{1,2,3,4,5,10,20\}$)  
for four cut-off times, respectively. 
In Table \ref{tbl:results20150401}, we present different parameter settings of each method that achieve the best HR@$k$ results for 
each $k$ value. 
For simplicity, in Table \ref{tbl:results20141001}, \ref{tbl:results20140401} and \ref{tbl:results20131001},  
we only present the parameter setting of each method that achieve the best HR@$5$. 
%
We also present the improvement of \logmean over the second best results in terms of all HR metrics in the tables. 
In addition, we add the experimental results for the \logmean model when $m_s$ equals to the number of all previous 
search terms in the current search session for each patient  
(i.e., the row for \logmean with parameter ``all" in Table~\ref{tbl:results20150401}). 
%

\subsubsection{Overall Performance on Cut-off Time 04/01/2015}
\label{sec:results:overall:20150401}

\begin{table*}[ht!]
\small
  \caption{Performance Comparison with Cut-off Time 04/01/2015}
  \centering
  \label{tbl:results20150401}
  \begin{threeparttable}
      \begin{tabular}{
	@{\hspace{10pt}}l@{\hspace{10pt}}
        @{\hspace{7pt}}r@{\hspace{10pt}}
	@{\hspace{7pt}}r@{\hspace{10pt}}
        @{\hspace{7pt}}r@{\hspace{7pt}}
	@{\hspace{0pt}}l@{\hspace{7pt}}
	@{\hspace{10pt}}r@{\hspace{10pt}}          
	@{\hspace{10pt}}r@{\hspace{10pt}}
	@{\hspace{10pt}}r@{\hspace{10pt}}
	@{\hspace{10pt}}r@{\hspace{10pt}}
        @{\hspace{10pt}}r@{\hspace{10pt}}
        @{\hspace{7pt}}r@{\hspace{10pt}}
        @{\hspace{7pt}}r@{\hspace{10pt}}
	}
        \toprule
        method      & \multicolumn{3}{c}{parameters}  &  & HR@1  & HR@2 & HR@3 & HR@4 & HR@5 & HR@10 & HR@20 \\ 

        \midrule

\multirow{3}{*}{\HGN} 
{} & $5$ & $1$ & - & & \underline{0.0798} & \underline{0.1063} & 0.1247 & 0.1329 & 0.1472 & 0.1963 & 0.3129 \\
{} & $20$ & $3$ &  - & & 0.0654 & 0.1002 & \underline{0.1309} & \underline{0.1391} & \underline{0.1616} & \underline{0.2188} & 0.3395 \\
{} & $10$ & $2$ & -  & & 0.0593 & 0.0941 & 0.1125 & 0.1309 & 0.1554 & 0.2025 & \underline{0.3476} \\

\midrule

\multirow{6}{*}{\HAM} 
 & $9$ & $2$  & $1$  && {0.0675} &  {0.0941} &  {0.1145} &  {0.1391} &  \underline{0.1616} &  {0.1943} & {0.3211}  \\ 
 & $30$ & $3$ & $3$ & &  \underline{0.0818} & 0.0941 & 0.1022 & 0.1043 & 0.1125 & 0.1963 & 0.3190 \\
 & $30$ & $3$ & $2$ & &  0.0736 & \underline{0.1084} & 0.1247 & 0.1350 & 0.1411 & 0.1738 & 0.3292 \\
{} & $15$ & $2$ & $3$ & &  0.0736 & {0.0961} & \underline{0.1288} & \underline{0.1431} & 0.1472 & 0.1861 & 0.3374 \\
{} & $30$ & $3$ & $1$ & &  0.0613 & 0.0941 & 0.1125 & 0.1309 & 0.1411 & \underline{0.2168} & 0.3354 \\
{} & $4$ & $3$ & $2$ & &  0.0716 & 0.0961 & 0.1145 & 0.1288 & 0.1391 & 0.1902 & \underline{0.3517} \\

\midrule

\multirow{4}{*}{\baseline} 
 & $0.1$ & $0.1$ & $0.9$    &&  {\underline{\bf{0.0859}}} & \underline{0.1309} & \underline{0.1595} & 0.1861 & 0.2045 & 0.2638 & 0.3722  \\
 & $0.1$ & $0.1$ & $0.7$   &&  0.0736 & 0.1247 & 0.1575 & \underline{0.1881} & \underline{0.2106} & 0.2945 & 0.3742  \\
 & $0.1$ & $0.1$ & $0.5$   &&  0.0736 & 0.1186 & 0.1493 & 0.1779 & 0.1984 & \underline{0.3006} & 0.3763  \\
 & $0.1$ & $0.1$ & $0.3$   &&  0.0716 & 0.1145 & 0.1472 & 0.1800 & 0.1984 & 0.2822 & \underline{0.3804}  \\

\midrule

\personlized                  &  - &  - & -  & & \underline{0.0734} & \underline{0.1193} & \underline{0.1835} & \underline{0.2110} & \underline{\bf{0.2294}} & \underline{0.2844} & \underline{0.3211} \\ 

        \midrule
\multirow{3}{*}{\encounter} 
                      & $64$ & $0.01$ & -  &  & 0.0642 & 0.1284 & \textbf{\underline{0.2018}} & \textbf{\underline{0.2202}} & \textbf{\underline{0.2294}} & 0.3211 & *\underline{\textbf{0.4312}}     \\
                      & $32$ & $0.05$ &  - &  & 0.0642 & 0.1284 & 0.1835 & 0.2018 & 0.2202 & \underline{\bf{0.3303}} & 0.4128     \\
                      & $32$ & $0.01$ &  - &  & \underline{0.0734} & \textbf{\underline{0.1376}} & 0.1927 & 0.2018 & 0.2202 & 0.3211 & *\underline{\textbf{0.4312}}     \\

\midrule

\multirow{4}{*}{\logmean} 
&  6  & 2 & 0.2  && 0.0642 & *\underline{\textbf{0.1468}} &*\underline{\textbf{0.2110}} & 0.2294 & *\underline{\textbf{0.2569}} & 0.3394 & 0.3945 \\ 
&  6  & 2 & 0.0  && 0.0550 & 0.1376 & 0.1927 & *\underline{\textbf{0.2477}} & 0.2477 & *\underline{\textbf{0.3486}} & 0.3945 \\ 
& 4  & 1 & 1.0  && *\underline{\textbf{0.0917}} & 0.1376 & 0.1743 & 0.2294 & *\underline{\textbf{0.2569}} & 0.3211 & \underline{\bf{0.4220}} \\ 
   & all & 2 & 0.8  && 0.0734 & 0.1376 & 0.1835 & 0.2110 & 0.2477 & 0.3303 & 0.3945 \\

\midrule

Improvement    &  &   &   && 6.75\% & 6.69\% & 4.56\% & 12.49\% & 11.99\% & 5.54\% & -2.13\% \\

        \bottomrule
      \end{tabular}
      \begin{tablenotes}[normal,flushleft]
      \begin{scriptsize}
      \item
          The best performance under each metric over all the methods is \textbf{bold} with *.
          The second best performance for under each metric over all the methods is \textbf{bold}. 
          The best performance within each method under each metric is \underline{underlined}.
          %
          The three parameters for \baseline are patient similarity threshold, term similarity threshold and weighting parameter~\cite{fan2019improving}; 
          the two parameters for \HGN are the number of previous purchases/ratings that are used for recommendation and the 
          number of next purchases/ratings that are recommended for~\cite{ma2019hierarchical}; 
          the three parameters for \HAM are the number of items of high-order sequential association, 
           the number of items of low-order sequential association, and the 
          number of next purchases/ratings that are recommended for~\cite{Peng2020HAM}; 
          the two parameters for \encounter are dimension of latent features and regularization weight; 
          the three parameters for \logmean are the number of previous search terms that are used for recommendation, the 
          number of previous encounters that are used for recommendation and the weighting parameter (i.e., $\alpha$). 
          We didn't present latent dimension $k$ in the table since it is equal to 32 for all the \logmean results. 
          %
          The value ``all" for parameter $m_s$ indicating considering all previous search terms for recommendation. 
         \par
      \end{scriptsize}
      \end{tablenotes}

  \end{threeparttable}
  
  \end{table*}



Table \ref{tbl:results20150401} shows that the \logmean model outperforms all the baseline methods 
in terms of HR@$k$ with $k \in \{1,2,3,4,5,10\}$, and achieves the second best 
result in terms of HR@$20$. 
%
\logmean achieves HR@$1$=$0.0917$, meaning that about 9\% of the patients have their ground-truth search terms correctly 
recommended at the very top. 
Although this HR@$1$ value is not high by itself, 
it is actually substantially higher than random guessing, which has an expected HR@$1$=$\frac{1}{1,915}$=$5.22e^{-4}$
(1,915 search terms in training set for cut-off time 04/01/2015 as in Table~\ref{tbl:dataset_cutoff}). 
That is, \logmean is 175-fold better than random guessing on HR@$1$. 
\logmean achieves HR@$5$=$0.2569$, meaning that about 26\% of the patients have their ground-truth search terms 
among top-5 recommendations. This is also substantially higher than random guessing, which has an expected 
HR@$5$=$5.22e^{-4}\times 5$=$2.6e^{-3}$. That is, \logmean is 98-fold better than random guessing on HR@$5$. 
%

%
The second best method is \encounter as it achieves the second best results  on 
HR@$2$, HR@$3$, HR@$4$, HR@$5$ and HR@$10$, and the best results on HR@$20$. 
The difference between \logmean and \encounter is that \logmean uses the most recent search terms and the most recent encounters to 
recommend the next search term, whereas \encounter uses all the previous encounters to recommend the next search term. 
However, as Table \ref{tbl:dataset_cutoff} shows, each patient has 167.88 encounters on average. 
Since previous encounters that occurred long time ago may not contain the information that the clinicians want to search at the time of recommendation, 
therefore, \encounter may generate inaccurate search term recommendations using all previous encounters. 
%
%
%
%
%
%

%
\personlized is slightly worse than \encounter and \logmean. This is probably because \personlized uses the most popular search terms  
in a patient's entire searching history for recommendation, 
and the popular terms that occurred long time ago may not be of interest to clinicians at the time of recommendation, 
given the progression of the patient's health condition. 
%
%
\baseline has slightly worse performance than \personlized. This might be due to the data sparsity issue. 
As shown in Table \ref{tbl:dataset_cutoff}, the dataset contains 
2,627 patients and 1,915 unique search terms in the training set. However, each patient has only 8.22 search terms on average. 
The data sparsity issue may cause \baseline to learn inaccurate patient-patient similarities and term-term similarities, both of 
which can lead to inaccurate search term recommendations. 

%
\HAM and \HGN have the worst performance in this experiment. 
%
%
In the RS settings, \HGN assumes that each item contributes differently to  next item recommendations, and  therefore 
learns  importance weights for  items using gating mechanisms. 
However, the data sparsity issue in our dataset may cause \HGN to learn inaccurate weights for items, and thus 
leads to poor recommendation performance. 
%
%
On the other hand, \HAM generates recommendations using three factors: 
users' long-term preferences modeled from all previous items, 
high-order association patterns modeled from a number of most recent previous items  
and low-order association patterns modeled from few most recent previous items. 
%
Our dataset may not contain the long-term preferences and the association patterns, and 
thus \HAM may generate less meaningful recommendations and thus have 
poor recommendation performance. 
%

Table \ref{tbl:results20150401} also shows that \logmean consistently improves over the second best method on all the 
evaluation metrics except  HR@$20$. In real applications, we prefer good performance with small values $k$ in HR@$k$, 
indicating that the correct recommendations are on very top. 
Therefore, good performance on HR@$k$, $k<20 $ is of more significance 
than that on HR@$20$.
In addition, \logmean
has relatively high improvement (more than $10\%$) over the second best results in terms of 
HR@$4$ and HR@$5$. This indicates that \logmean can push the most relevant search terms on very top of the recommendation list. 


\subsubsection{Overall Performance on Other Cut-off Times}
\label{sec:results:overall:other}

\begin{table*}[ht!]
\small
  \caption{Performance Comparison with Cut-off Time 10/01/2014}
  \centering
  \label{tbl:results20141001}
  \begin{threeparttable}
     \vspace{-5pt}
      \begin{tabular}{
	@{\hspace{8pt}}l@{\hspace{8pt}}
        @{\hspace{8pt}}r@{\hspace{8pt}}
	@{\hspace{8pt}}r@{\hspace{8pt}}
        @{\hspace{8pt}}r@{\hspace{8pt}}
	@{\hspace{8pt}}l@{\hspace{8pt}}
	@{\hspace{8pt}}r@{\hspace{8pt}}          
	@{\hspace{8pt}}r@{\hspace{8pt}}
	@{\hspace{8pt}}r@{\hspace{8pt}}
	@{\hspace{8pt}}r@{\hspace{8pt}}
        @{\hspace{8pt}}r@{\hspace{8pt}}
        @{\hspace{8pt}}r@{\hspace{8pt}}
        @{\hspace{8pt}}r@{\hspace{8pt}}
	}
        \toprule
        method      & \multicolumn{3}{c}{parameters}  &  & HR@1  & HR@2 & HR@3 & HR@4 & HR@5 & HR@10 & HR@20 \\ 

        \midrule

\HGN & $20$ & $3$  &  - && 0.0854 & 0.1250 & 0.1437 & 0.1604 & 0.1833 & 0.2354 & 0.3083\\ 
\HAM & $30$ & $1$  & $3$  && 0.0583 & 0.0854 & 0.1229 & 0.1521 & 0.1854 & 0.2396 & 0.3208 \\
\baseline  &   $0.1$ & $0.1$ & $0.1$ & & {0.0750} & {0.1146} & {0.1479} & {0.1708} & {0.1979} & {0.2729} & \textbf{0.3479} \\
\personlized                 &  - &  - &  - & &              *\textbf{0.1500}   & \textbf{0.1500} & \textbf{0.1500} & 0.1500 & 0.1500 & 0.1500 & 0.1500 \\   
\encounter                   & $32$ & $0.05$ & -  & &          \textbf{0.1000}   & \textbf{0.1500} & \textbf{0.1500} & \textbf{0.2000} & \textbf{0.2000} & \textbf{0.3000} & *\textbf{0.4000} \\ 
        \midrule

\logmean            &10   & 1 & 0.0  && *\textbf{0.1500} & *\textbf{0.2000} & *\textbf{0.2000} & *\textbf{0.2500} & *\textbf{0.2500} & *\textbf{0.3500} & *\textbf{0.4000} \\


\midrule

Improvement    &  &   &   && 50.00\% & 33.33\% & 33.33\% & 25.00\% & 25.00\% & 16.67\% & 14.98\% \\

        \bottomrule
      \end{tabular}

      \begin{tablenotes}[normal,flushleft]
      \begin{scriptsize}
      \item
         In this table,
         the best performance under each metric over all the methods is \textbf{bold} with *.
          The second best performance for under each metric over all the methods is \textbf{bold}. 
          The parameter columns of each method are corresponding to those in Table \ref{tbl:results20150401}. 
          %
          
                \par
              
      \end{scriptsize}
      \end{tablenotes}

  \end{threeparttable}

\end{table*}


\begin{table*}[!t]

\small
  \caption{Performance Comparison with Cut-off Time 04/01/2014}
  \centering
  \label{tbl:results20140401}
  \begin{threeparttable}
      \vspace{-5pt}
      \begin{tabular}{
	@{\hspace{8pt}}l@{\hspace{8pt}}
        @{\hspace{8pt}}r@{\hspace{8pt}}
	@{\hspace{8pt}}r@{\hspace{8pt}}
        @{\hspace{8pt}}r@{\hspace{8pt}}
	@{\hspace{8pt}}l@{\hspace{8pt}}
	@{\hspace{8pt}}r@{\hspace{8pt}}          
	@{\hspace{8pt}}r@{\hspace{8pt}}
	@{\hspace{8pt}}r@{\hspace{8pt}}
	@{\hspace{8pt}}r@{\hspace{8pt}}
        @{\hspace{8pt}}r@{\hspace{8pt}}
        @{\hspace{8pt}}r@{\hspace{8pt}}
        @{\hspace{8pt}}r@{\hspace{8pt}}
	}
        \toprule
        method      & \multicolumn{3}{c}{parameters}  &  & HR@1  & HR@2 & HR@3 & HR@4 & HR@5 & HR@10 & HR@20 \\ 

        \midrule

\HGN & $2$  & $3$  &  - && 0.1458 & 0.1852 & 0.2075 & 0.2281 & 0.2556 & 0.2916 & 0.3688 \\

\HAM & $25$ & $0$  & $3$  && 0.0755 & 0.1578 & 0.2007 & 0.2367 & 0.2590 & 0.3002 & 0.3585 \\

\baseline  &   $0.1$ & $0.1$ & $0.7$ & & {0.1407} & {0.2024} & {0.2264} & {0.2436} & {0.2676} & \textbf{0.3276} & {0.3928} \\ 

\personlized                 &  - &  - &  - & &              0.2302   & 0.2374 & 0.2734  & 0.2950  & 0.3094  & 0.3094  & 0.3094 \\   
\encounter                       & 32 & 0.05 & -  & &          \textbf{0.3165}   & \textbf{0.3453} & \textbf{0.3741}  & \textbf{0.3957}  & \textbf{0.4029}  & *\textbf{0.4460}  & \textbf{0.4820} \\ 

        \midrule

\logmean &  7   & 1 & 0.4  && *\textbf{0.3237} & *\textbf{0.3597} & *\textbf{0.4101} & *\textbf{0.4245} & *\textbf{0.4317} & *\textbf{0.4460} & *\textbf{0.5324} \\

\midrule

Improvement    &  &   &   && 7.20\% & 4.17\% & 10.42\% & 7.28\% & 7.15\% & 36.14\% & 10.46\% \\

        \bottomrule
      \end{tabular}
      
      \begin{tablenotes}[normal,flushleft]
      \begin{scriptsize}
      \item
         In this table,
         the best performance under each metric over all the methods is \textbf{bold} with *.
          The second best performance for under each metric over all the methods is \textbf{bold}. 
          The parameter columns of each method are corresponding to those in Table \ref{tbl:results20150401}. 
                \par
              
      \end{scriptsize}
      \end{tablenotes}

  \end{threeparttable}
\end{table*}

\begin{table*}[!h]

\small
  \caption{Performance Comparison with Cut-off Time 10/01/2013}
  \centering
  \label{tbl:results20131001}
  \begin{threeparttable}
    \vspace{-5pt} 
     \begin{tabular}{
	@{\hspace{8pt}}l@{\hspace{8pt}}
        @{\hspace{8pt}}r@{\hspace{8pt}}
	@{\hspace{8pt}}r@{\hspace{8pt}}
        @{\hspace{8pt}}r@{\hspace{8pt}}
	@{\hspace{8pt}}l@{\hspace{8pt}}
	@{\hspace{8pt}}r@{\hspace{8pt}}          
	@{\hspace{8pt}}r@{\hspace{8pt}}
	@{\hspace{8pt}}r@{\hspace{8pt}}
	@{\hspace{8pt}}r@{\hspace{8pt}}
        @{\hspace{8pt}}r@{\hspace{8pt}}
        @{\hspace{8pt}}r@{\hspace{8pt}}
        @{\hspace{8pt}}r@{\hspace{8pt}}
	}
        \toprule
        method      & \multicolumn{3}{c}{parameters}  &  & HR@1  & HR@2 & HR@3 & HR@4 & HR@5 & HR@10 & HR@20 \\ 

        \midrule

\HGN & $25$ & $3$  & -  && 0.0526 & 0.0643 & {0.0994} & 0.1053 & 0.1404 & 0.1637 & 0.2339 \\

\HAM & $10$ & $3$  & $3$  && 0.0468 & \textbf{0.0994} & \textbf{0.1637} & 0.1696 & \textbf{0.1988} & \textbf{0.2456} & *\textbf{0.3216} \\
\baseline  &   $0.1$ & $0.1$ & $0.7$ & &  {0.0175} & {0.0234} & {0.0292} & {0.0351} & {0.0468} & {0.0760} & {0.1345} \\

\personlized                 &  - &  - &  - & &        *\textbf{0.1818}   &  *\textbf{0.1818} & *\textbf{0.1818} & \textbf{0.1818} & 0.1818 & 0.1818 & 0.1818\\   
\encounter                   & 32 & 0.05 &  - & &        \textbf{0.0909}   &  {0.0909} & 0.0909 & 0.0909 & *\textbf{0.2727} & *\textbf{0.2727} & \textbf{0.2727} \\

        \midrule

\logmean  &   10 & 1 & 0.0  && \textbf{0.0909} & {0.0909} & *\textbf{0.1818} & *\textbf{0.2727} & *\textbf{0.2727} & *\textbf{0.2727} & \textbf{0.2727}  \\

\midrule

Improvement    &  &   &   && -50.00\% & -50.00\% & 11.06\% & 50.00\% & 37.17\% & 11.03\% & -15.21\% \\

        \bottomrule
      \end{tabular}
      \begin{tablenotes}[normal,flushleft]
      \begin{scriptsize}
      \item
         In this table,
         the best performance under each metric over all the methods is \textbf{bold} with *.
          The second best performance for under each metric over all the methods is \textbf{bold}. 
          The parameter columns of each method are corresponding to those in Table \ref{tbl:results20150401}. 
                \par
              
      \end{scriptsize}
      \end{tablenotes}
 
  \end{threeparttable}

\end{table*}


Table \ref{tbl:results20141001} and Table \ref{tbl:results20140401} presents the performance of the different methods 
on cut-off time 10/01/2014 and 04/01/2014, respectively. 
They show similar trends as those in 
Table \ref{tbl:results20150401}, and \logmean outperforms the baseline methods in terms of all HR metrics. 
%
%
Further comparing Table \ref{tbl:results20150401}, Table \ref{tbl:results20141001} and 
Table \ref{tbl:results20140401}, we  
notice that all the methods tend to have higher HR results in Table  \ref{tbl:results20140401} (cut-off time 04/01/2014) than those in 
Table \ref{tbl:results20141001} (cut-off time 10/01/2014) and Table \ref{tbl:results20150401} (cut-off time 04/01/2015). 
This is probably because there are fewer search terms (i.e., 1,065) in the training set with cut-off time 
04/01/2014 compared to those with cut-off-time 10/01/2014 (i.e., 1,493) and those
with cut-off time 04/01/2015 (i.e., 1,915), as shown in Table \ref{tbl:dataset}. 
Note that in our experiments, the model will only recommend the search terms that have appeared in the training set, 
and will not recommend new search terms. 
Since generating recommendations from 1,065 search terms is 
more likely to hit the accurate term by chance
than generating recommendations from 1,915 search terms.  
Thus, 
\logmean is more likely to recommend search terms accurately with cut-off time 04/01/2014 
than with cut-off time 04/01/2015. 
%

%

%

%

Table \ref{tbl:results20131001} presents the performance of the different methods 
on cut-off time 10/01/2013. It shows that  \logmean achieves the best results  
in terms of HR@$3$, HR@$4$, HR@$5$ and HR@$10$, 
and achieves the second best results in terms of HR@$1$ and HR@$20$. 
In this experiment, \personlized achieves the best HR@$1$, HR@$2$ and HR@$3$ results. 
Furthermore, \HAM achieves the best result in terms of HR@$20$, and achieves the second best 
results in terms of HR@$2$, HR@$3$, HR@$5$ and HR@$10$. 
In this experiment, 
there are very few patients and search terms in the training set and testing set, as shown in Table \ref{tbl:dataset}. 
Due to the data sparsity, the representations of search terms and ICD 9 codes may not be well learned, and thus 
it may cause both \logmean and \encounter  to have relatively poor recommendation performance, and do not achieve 
significant improvement over the baseline methods. 
%
%
Table \ref{tbl:results20131001} also shows that \personlized has the same values for all the HR metrics. 
This is probably because there are sessions in which testing search terms do not appear in the training part 
of the same sessions. Therefore, \personlized is not able to recommend the accurate search terms for such sessions 
despite of the value of $k$.  

Further comparing Table \ref{tbl:results20150401}, \ref{tbl:results20141001}, \ref{tbl:results20140401} 
and \ref{tbl:results20131001},  
we notice that \logmean tends to significantly outperform the baseline methods when there is sufficient data 
for training, and would  
not perform very well when data is very sparse. 
Overall, for all three datasets,  \logmean consistently outperforms the baseline methods in terms of 
HR@$3$, HR@$4$, HR@$5$ and HR@$10$, and 
achieves improvement of at least 4.56\%, 7.28\%, 7.15\% and 
5.54\% over the second best method, respectively. 
\subsection{Parameter Study}
\label{sec:results:parameter}
%
%

For a parameter study, we evaluate the models with cut-off time 04/01/2015 
since there is sufficient training and testing data with this cut-off time.  
We choose the set of parameter as follows: $m_s$=$6$, $m_c$=$2$ and  $\alpha$=$0.2$, 
since \logmean 
achieves the best HR@$2$, HR@$3$ and HR@$5$ with this set of parameters. 
To conduct parameter study, we fix two out of three parameters and evaluate the model with 
different values of the third parameter. 
Table \ref{tbl:parameter-alpha}, \ref{tbl:parameter-lc} and \ref{tbl:parameter-ls} present the results
for parameter study on \logmean on $m_s$, $m_c$ and $\alpha$, respectively.  
%

\begin{table}[!t]
\centering
\small
  \caption{Parameter Study of \logmean on $m_s$ ($m_c$=$2$, $\alpha$=$0.2$)}
  \centering
  \label{tbl:parameter-ls}
  \begin{threeparttable}
      \begin{tabular}{
	@{\hspace{18pt}}l@{\hspace{18pt}}
	@{\hspace{18pt}}r@{\hspace{18pt}}
	@{\hspace{18pt}}r@{\hspace{18pt}}
	@{\hspace{18pt}}r@{\hspace{18pt}}
	}
        \toprule
        $m_s$  &  HR@1  &  HR@5 & HR@10  \\ 

        \midrule
        1 & 0.0642 & 0.2110 & \textbf{0.3486} \\
        2 & 0.0642 & 0.2477 & 0.3303 \\
        3 & 0.0642 & 0.2202 & 0.3394 \\
        4 & 0.0642 & 0.2477 & 0.3394 \\
        $[5,9]$ & 0.0642 & \textbf{0.2569} & 0.3394 \\
        10 & 0.0642 & 0.2477 & 0.3394 \\
        all & 0.0642 & 0.2202 & 0.3394 \\

        \bottomrule
      \end{tabular}
          \begin{tablenotes}[normal,flushleft]
          \begin{scriptsize}
          \item
              In this table, 
              the best performance for each metric is \textbf{bold}. 
%
              \par
          \end{scriptsize}
          \end{tablenotes}
  \end{threeparttable}

\end{table}

\begin{table}[!h]
\centering
\small
  \caption{Parameter Study of \logmean on $m_c$ ($m_s$=$6$, $\alpha$=$0.2$)}
  \vspace{-3pt}
  \centering
  \label{tbl:parameter-lc}
  \begin{threeparttable}
      \begin{tabular}{
	@{\hspace{18pt}}l@{\hspace{18pt}}
	@{\hspace{18pt}}r@{\hspace{18pt}}
	@{\hspace{18pt}}r@{\hspace{18pt}}
	@{\hspace{18pt}}r@{\hspace{18pt}}
	}
        \toprule
         $m_c$  &  HR@1  &  HR@5 & HR@10  \\ 

        \midrule
        1 & 0.0642 & 0.2110 & 0.2844 \\
        2 & 0.0642 & \textbf{0.2569} & \textbf{0.3394} \\
        3 & 0.0642 & 0.2385 & \textbf{0.3394} \\
        4 & 0.0642 & 0.2294 & 0.3303 \\

        \bottomrule
      \end{tabular}
          \begin{tablenotes}[normal,flushleft]
          \begin{scriptsize}
          \item
              In this table, 
              the best performance for each metric is \textbf{bold}. 
%
              \par
          \end{scriptsize}
          \end{tablenotes}
  \end{threeparttable}
\end{table}

\begin{table}[!h]
\centering


\small
  \caption{Parameter Study of \logmean on $\alpha$ ($m_s$=$6$, $m_c$=$2$)}
  \centering
  \label{tbl:parameter-alpha}
  \begin{threeparttable}
      \begin{tabular}{
	@{\hspace{18pt}}l@{\hspace{18pt}}
	@{\hspace{18pt}}r@{\hspace{18pt}}
	@{\hspace{18pt}}r@{\hspace{18pt}}
	@{\hspace{18pt}}r@{\hspace{18pt}}
	}
        \toprule
        $\alpha$ &  HR@1  &  HR@5 & HR@10  \\ 

        \midrule

         0.0   & 0.0550 & 0.2477 & \textbf{0.3486} \\
         0.2   & 0.0642 & 0.\textbf{2569} & 0.3394 \\
         0.4   & 0.0642 & 0.\textbf{2569} & 0.3303 \\
         0.6   & 0.0734 & 0.2477 & 0.3211 \\
         0.8   & 0.0734 & 0.\textbf{2569} & 0.3394 \\
         1.0   & \textbf{0.0917} & 0.2294 & 0.3119 \\

        \bottomrule

      \end{tabular}
          \begin{tablenotes}[normal,flushleft]
          \begin{scriptsize}
          \item
              In this table, 
              the best performance for each metric is \textbf{bold}. 
              \par
          \end{scriptsize}
          \end{tablenotes}
  \end{threeparttable}
\end{table}

Table \ref{tbl:parameter-ls} shows that \logmean achieves the best HR@$5$ result with $m_s \in [5,9]$ 
and the value of HR@$5$ decreases when $m_s$ increases or decreases from the range. 
Recall that $m_s$ is the number of previous search terms that are used to recommend the next search term.  
The result here indicates that 
too few previous search terms may not provide sufficient information  
and too many pervious search terms may provide irrelevant information for recommendation. 
Table \ref{tbl:parameter-ls}  also shows that \logmean achieves  the same HR@$1$ result with different 
$m_s$ values. This indicates that increasing or decreasing $m_s$ values may not be able to push the most relevant 
search terms to the very top of the recommendation list, but can rank more relevant search terms on top 
since HR@$5$ is improved when choosing a proper $m_s$. 
%
%

%
Table \ref{tbl:parameter-lc}  shows that \logmean achieves  the same HR@$1$ result with different 
$m_c$ values, achieves the best HR@$5$ with $m_c$$=$$2$ and achieves the best HR@$5$ 
with $m_c$$=$$2$ and $3$. 
Recall that $m_c$ is the number of previous encounters that are used to recommend the next search term. 
Table \ref{tbl:parameter-lc}   also shows that both HR@$5$ and HR@$10$ 
decrease as $m_c$ increases or decreases from 2 (or 3). 
This indicates that changing $m_c$ values may not help \logmean to push the most relevant search terms 
on the very top of the recommendation list, but when $m_c$ is set to a small number (e.g., $m_c$=$2$), 
\logmean can achieve its best results in terms of HR@$5$ and HR@$10$. 


Table \ref{tbl:parameter-alpha} shows that when $\alpha$ (Equation \ref{eqn:recommendation}) increases, HR@$1$ increases, 
and HR@$10$ tends to decrease. 
Furthermore, \logmean achieves the best HR@$5$ results with $\alpha$ equal to $0.2$, $0.4$, $0.8$. 
Recall that with $\alpha$ set to $1.0$, \logmean only uses previous search terms to generate recommendations, and 
with  $\alpha$ as $0$, \logmean only uses previous encounters for recommendation. 
This indicates that having more weights on previous search terms can increase HR@$1$ performance, 
but hurt HR@$10$ performance, 
whereas adding more weights to previous encounters can hurt HR@$1$ performance, but improve HR@$10$ performance.
This is probably because using previous search terms can push the most relevant 
search terms to the very top of the recommendation list, thus, \logmean can achieve 
the best HR@$1$ result with $m_s$ equal to $1.0$. 
However, only using previous search terms overlooks encounters' information, which may be highly relevant 
to clinicians' searches. 
Therefore,  \logmean  can push more relevant search terms on top, and achieves the best HR@$10$ with $m_s$ equal to $0$. 
When using both previous search terms and previous encounters, \logmean is able to achieves the best HR@$5$ results.

\subsection{Analysis on Length of Search Sessions}
\label{sec:results:session_length}

\begin{table}[h!]
\small
\centering
\captionof{table}{Statistics on Lengths of Sessions}
  \label{tbl:session_length}
      \begin{tabular}{
        @{\hspace{6pt}}l@{\hspace{6pt}}
        @{\hspace{6pt}}r@{\hspace{6pt}}
        @{\hspace{6pt}}r@{\hspace{6pt}}
        @{\hspace{6pt}}r@{\hspace{6pt}}
        }
        \toprule
        length of sessions & min   & max & mean \\
        \midrule
        top 20\% shortest sessions       & 2 & 2 & 2 \\
        top 20\%-40\% shortest sessions  & 3 & 5 & 3.95 \\
        mid 40\%-60\% sessions           & 6 & 8 & 6.76 \\
        top 20\%-40\% longest sessions   & 9 & 24 & 15.04 \\
        top 20\% longest sessions        & 26 & 202 & 62.75 \\
        \bottomrule
      \end{tabular}
\end{table}

\begin{figure*}
\centering
\begin{subfigure}{.4\textwidth}
\centering
\includegraphics[width=\linewidth]{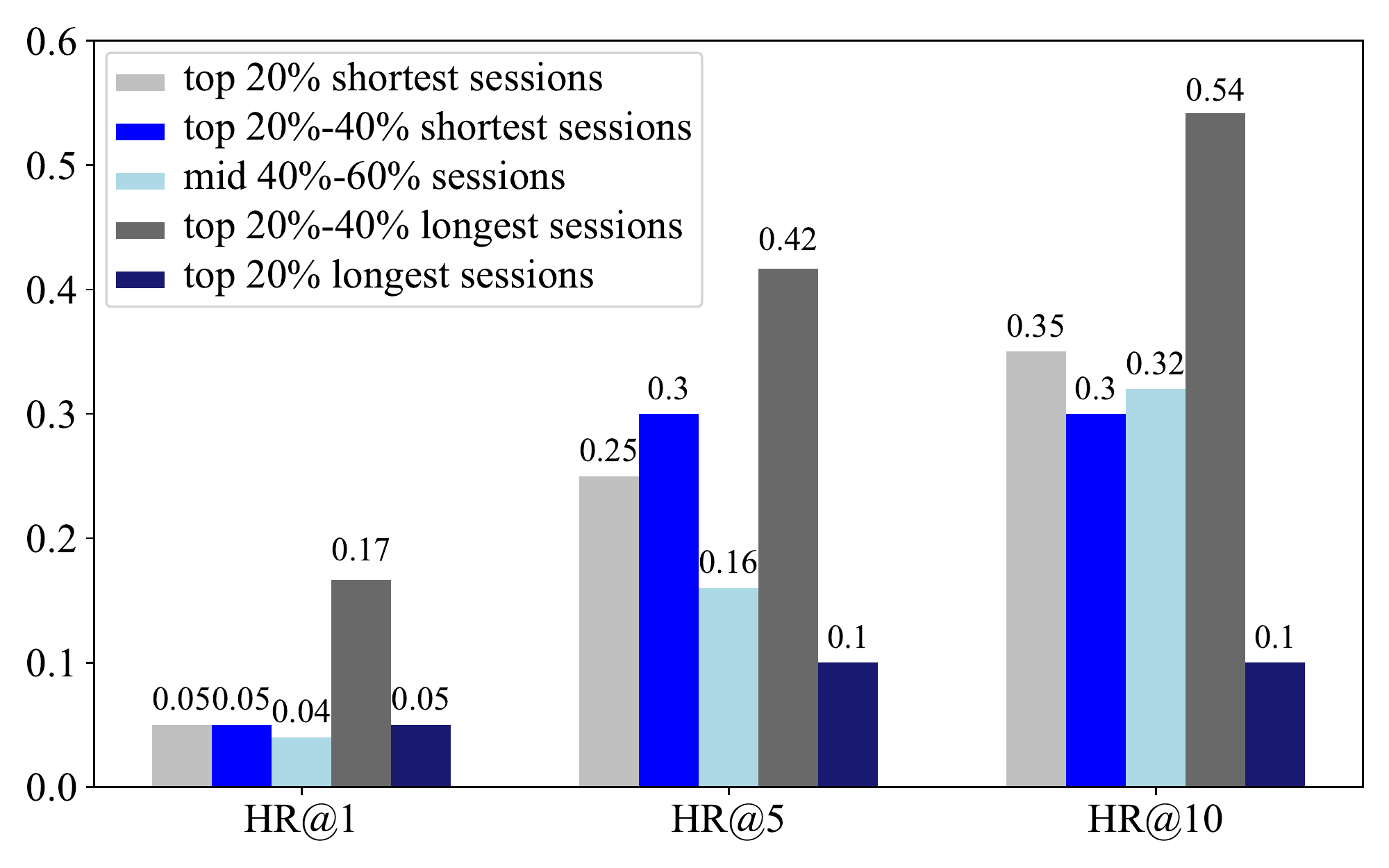}
\vspace{-20pt}
\caption{Parameter Settings: $m_s$=``all", $m_c$=$2$, $\alpha$=$0.8$}
\label{fig:session_length_all_2_0.8}
\end{subfigure}
\hspace{5mm}
\begin{subfigure}{.4\textwidth}
\centering
\includegraphics[width=\linewidth]{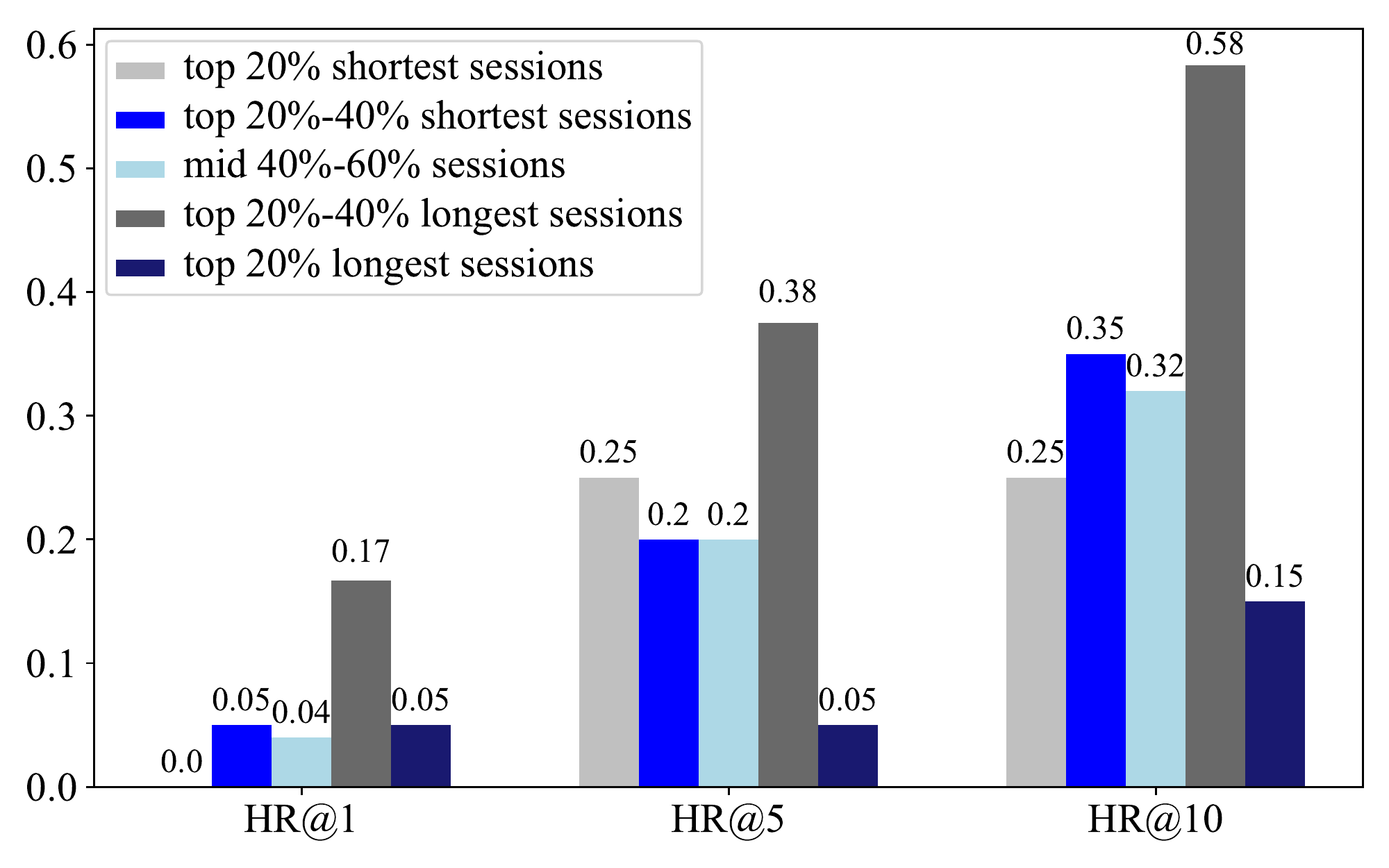}
\vspace{-20pt}
\caption{Parameter Settings: $m_s$=``all", $m_c$=$2$, $\alpha$=$0.2$}
\label{fig:session_length_all_2_0.2}
\end{subfigure}\par
\begin{subfigure}{.4\textwidth}
\centering
\includegraphics[width=\linewidth]{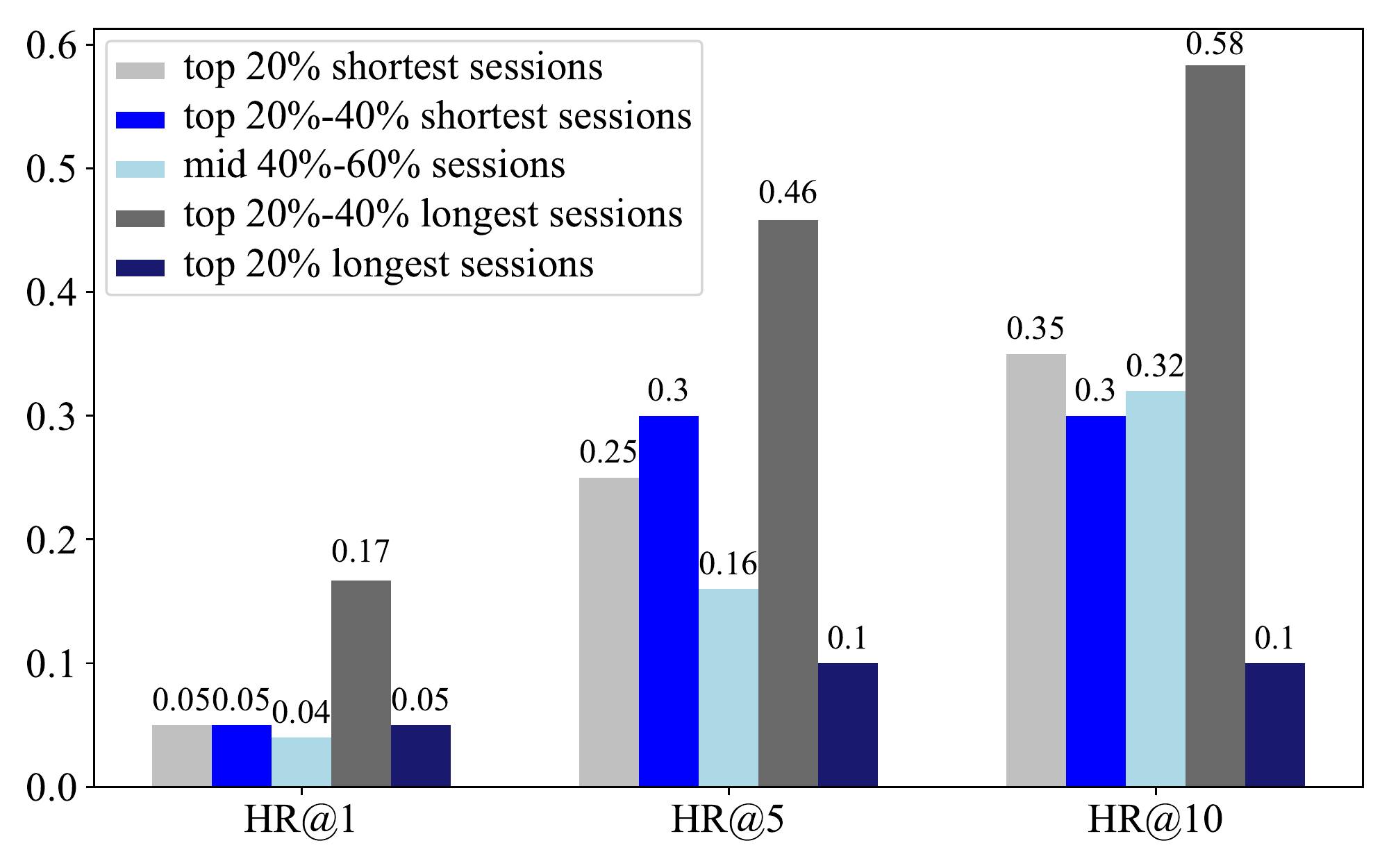}
\vspace{-20pt}
\caption{Parameter Settings: $m_s$=6, $m_c$=$2$, $\alpha$=$0.8$}
\label{fig:session_length_6_2_0.8}
\end{subfigure}
\hspace{5mm}
\begin{subfigure}{.4\textwidth}
\centering
\includegraphics[width=\linewidth]{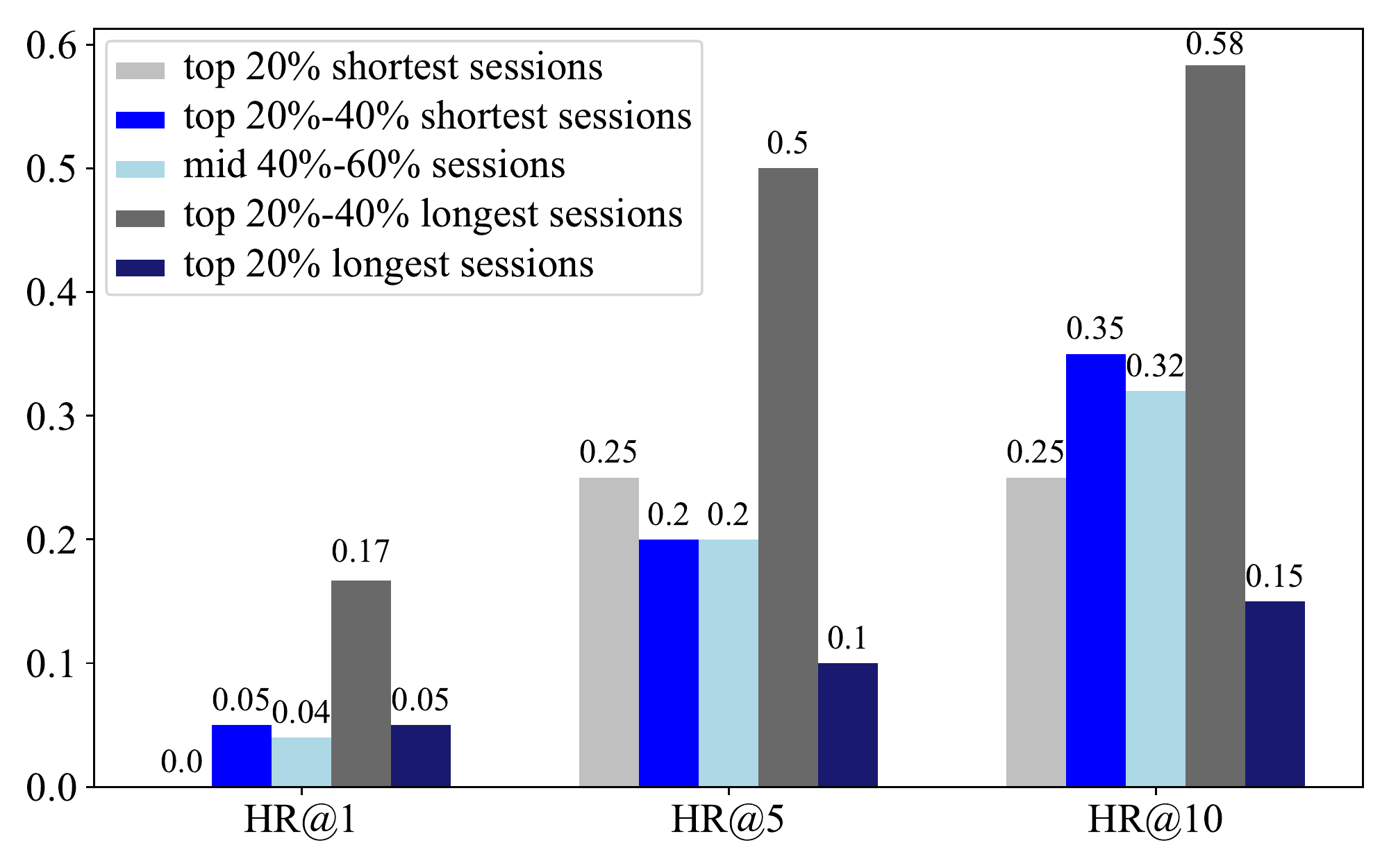}
\vspace{-20pt}
\caption{Parameter Settings: $m_s$=6, $m_c$=$2$, $\alpha$=$0.2$}
\label{fig:session_length_6_2_0.2}
\end{subfigure}\hfill

\caption{Performance Comparison for Sessions with Different Lengths}
\label{fig:session_length}

\end{figure*}



%
%
We evaluate \logmean on sessions of different sequence lengths 
in order to better understand the  influence of the session lengths and information content on the 
recommendation performance. 
Specifically, we first divide the sessions in the testing set into five groups based on the session lengths. 
Table \ref{tbl:session_length} presents the statistics of the lengths of search sessions in our experiments. 
For each group, we present the min length, max length and average length of the sessions. 
In this experiment, we include all the previous searches for the next search term recommendation, 
therefore, we set $m_s$ as ``all" in \logmean. 
Given $m_s$ as ``all", $m_c$=$2$ and $\alpha$=$0.8$ enable the best 
performance of \logmean, as shown in Table \ref{tbl:results20150401}. 
Thus, we select $m_s$=``all", $m_c$=$2$ and $\alpha$=$0.8$ as the parameters in our experiment. 
In addition, $m_s$=6, $m_c$=$2$ and $\alpha$=$0.2$ in \logmean achieve the best HR@2, 
HR@3 and HR@5 results, as shown in Table \ref{tbl:results20150401}. 
For comparison purposes, we also add this set of parameters in our experiment. 
Finally, in order to understand the effects of parameter $\alpha$, 
we add another two parameter settings for \logmean including $m_s$=``all", $m_c$=$2$, $\alpha$=$0.2$; and 
$m_s$=6, $m_c$=$2$, $\alpha$=$0.8$, that is, with $\alpha$ varying and the other two fixed from the above two parameter
settings, respectively.
%
%
%
We present the results of these four parameter settings in Figure \ref{fig:session_length_all_2_0.8}, 
\ref{fig:session_length_all_2_0.2}, \ref{fig:session_length_6_2_0.8} and \ref{fig:session_length_6_2_0.2}, 
respectively. 
%
%
%

Figure \ref{fig:session_length_all_2_0.8}, 
\ref{fig:session_length_all_2_0.2}, \ref{fig:session_length_6_2_0.8} and \ref{fig:session_length_6_2_0.2} 
show the results in terms of HR@$1$, HR@$5$ and HR@$10$ 
for the five groups of search sessions. 
%
The four figures show similar trends, that is, 
top-20\%-40\% longest sessions achieve the best performance, 
 %
and the top-20\% longest sessions achieve the worst. 
This indicates that previous search terms from long time ago  may not represent 
the information that the clinicians intend to search at the time of recommendation.  
%
Furthermore, groups with top-20\% shortest sessions, top-20\%-40\% shortest sessions  
and mid-40\%-60\% sessions achieve similar HR@$1$ and HR@$10$, and are all much 
worse than those with top-20\%-40\% longest sessions. 
This indicates that  too few search terms may not contain sufficient information to produce  
accurate search term recommendations, while the top-20\%-40\% longest sessions, 
that is, the sessions with length from 9 to 24 (Table~\ref{tbl:session_length}), can achieve accurate recommendations 
with adequate search term information. 
%
%

Moreover, \logmean with $m_s$ as ``all" and $m_s$=$6$ show a similar trend. 
Specifically, for the sessions with length less or equal to $6$ 
(i.e., top-20\% shortest sessions and top-20\%-40\% shortest sessions), \logmean with the 
two parameter settings  have identical results. 
This is because when $m_s$ is larger than the number of all previous search terms in the current session, 
we will use all the search terms to recommend the next search term. 
In this case, $m_s$=$6$ will have the same results as $m_s$=``all". 
For the sessions of mid-40\%-60\% lengths, given that their average length is 6.76 as shown in 
Table \ref{fig:session_length}, 
the 6 most recent searches constitute 88.76\% of a search session on average. 
Thus, using the 6 most recent searches in a session shows similar trends 
as using all previous searches for these sessions.
For the top-20\%-40\% longest sessions, the average length is 15.04 as shown in Table \ref{fig:session_length}, 
and  the 6 most recent searches constitute 39.90\% of a search session on average.  
Furthermore, comparing Figure~\ref{fig:session_length_all_2_0.8} and \ref{fig:session_length_6_2_0.8}, 
Figure \ref{fig:session_length_all_2_0.2} and \ref{fig:session_length_6_2_0.2}, we notice that 
using previous 6 searches achieves significant improvement 
compared to using all previous searches in term of HR@$5$ for top-20\%-40\% longest sessions.
This indicates that although using the most recent 6 searches discards a lot of information in a session, 
the most recent, pertinent information still retains in these sessions and thus enables  
superior recommendation performance of \logmean.
%
%
%
%
%
For top-20\% longest sessions, the average length is 62.75 as shown in Table \ref{fig:session_length}. 
For these sessions,  the most recent 6 searches on average constitute 9.56\% of a search session. 
Therefore, using the most recent 6 searches for recommendation may lose too much information 
and may have equally bad recommendation results as using all the previous searches, which, on the other hand,  
may contain too much irrelevant information. 
%

\subsection{Case Study}
\label{sec:results:casestudy}

\begin{table}[!h]
\centering
\small
  \caption{Testing Search Terms and Frequencies}
  \centering
  \label{tbl:case-study-freq-terms}
  \begin{threeparttable}
      \begin{tabular}{rr|rr
	}
        \toprule
        
       \multicolumn{2}{c|}{Top-$5$ Hit Group} & \multicolumn{2}{c}{Top-$5$ Miss Group} \\
    \midrule
         Term & Freq (\%) & Term & Freq (\%)   \\ 

        \midrule
hiv & 2,144 (33.33)		&			cbc & 542 (8.33)\\
a1c & 854 (8.33)			&			troponin & 371 (8.33)\\
creatinine & 709 (8.33)	&			ct & 224 (8.33)\\
weight & 706 	(8.33)	&			neurol & 119 (8.33)\\
ekg & 634 	(8.33)	&			operation & 101 (8.33)\\
cbc & 542 	(8.33)	&			culture \& blood & 79 (8.33)\\
bmp & 475 	(8.33)	&			hep & 51 (8.33)\\
inr & 431 		(8.33)	&			cytol & 45 (8.33)\\
pap & 279 	(8.33)	&			prealb & 40 (8.33)\\
	&			&			tspot & 31 (8.33)\\
	&			&			aldo & 14 (8.33)\\
	&			&			form & 5 (8.33)\\
   
        \bottomrule
      \end{tabular}
      \begin{tablenotes}[normal,flushleft]
      \begin{scriptsize}
      \item
      In this table, 
      Top-$5$ Hit Group is the group in which testing search terms are among top-5 recommendations;
      Top-$5$ Miss Group is the group in which testing terms are not among top-5 recommendations; 
      ``Freq" indicates how many times each  term is  searched on all the patients;
      ``\%" is the perentage of search sessions that have a corresponding testing search term among 
      all the search sessions in each group. 
          \par
      \end{scriptsize}
      \end{tablenotes}
       \end{threeparttable}
\end{table}

We present a case study based on the results in Section~\ref{sec:results:session_length}. 
Specifically, we extract top-20\%-40\% longest sessions (as shown in Table \ref{fig:session_length}) 
and divide the sessions into two groups: top-5 hit group and top-5 miss group, 
based on whether their testing search terms are correctly  recommended among top 5 or not. 
%
%
It turns out that each group has 12 sessions. 
We examine the testing search terms of each group and investigate the following two questions: 
(1) is the testing search term frequently searched? and 
(2) has the testing search term appeared earlier in a same session?
%
Table \ref{tbl:case-study-freq-terms} presents the testing search terms and 
how many times they are searched on all patients 
in our dataset (i.e., including training set and testing set). 
%
Table~{\ref{tbl:case-study-freq-terms}}  also shows the percentage of search sessions that have a 
corresponding testing search term. 
Table \ref{tbl:case-study-freq-terms} shows that for 
top-$5$ hit group, all the search terms in the testing set are frequently searched, and are searched for at least 200 times. 
However, for top-$5$ miss group, only 3 search terms in the testing set are searched for over 200 times, and 7 search terms 
are searched for less than 100 times. 
This shows that if the testing search term is frequently searched, the model is more likely to generate accurate recommendations 
for this search term. 
On the other hand, if the testing search term is infrequently searched, it is hard for the model to generate accurate recommendations. 
%
Moreover, in top-$5$ hit group, 75.00\% of the sessions have their testing search terms also appear earlier in the same sessions, 
whereas in top-$5$ miss group, only 41.67\% of the sessions have the testing search terms in the earlier part of the same sessions. 
%
This indicates that for each session, if the testing search term has appeared earlier in the same session, 
\logmean is much likely to generate accurate recommendations for this search term. 

\section{Conclusion}
\label{sec:conclusion}

In this manuscript, we developed a model named 
Hybrid Collaborative Filtering model using encounter information for search term 
recommendation  (\logmean) for clinicians. 
The \logmean model generates recommendations for the next search using two factors: previous search terms 
and previous encounters. 
Specifically, \logmean recommends the terms that have high co-occurrence frequencies with the most recent ICD  codes  
and are highly relevant to the most recent search terms for the next search. 
We conduct comprehensive experiments on different datasets to compare the proposed model with the state-of-the-art 
baseline methods. 
The experimental results demonstrate that the proposed model can outperform all the 
baseline methods with improvement of at least 4.56\%, 7.28\%, 7.15\% and 
5.54\% over the second best results 
 in terms of  HR@$3$, HR@$4$ , HR@$5$ and HR@$10$, respectively, on different datasets with different sparsities. 
 %
We have also conducted comprehensive parameter studies to analyze the impact of the two factors.  
The experimental results show that when using the most recent previous 5 to 9 search terms and 
the most recent previous two encounters, with the weight of previous search term factor between 0.2 and 0.8, 
\logmean is most likely to generate accurate search term recommendations. 
Furthermore, 
we evaluated the \logmean model on individual session groups divided by session length. 
The experimental results show that the search sessions with length from 9 to 24 are 
more likely to have better performance than a shorter or longer session. 
Finally, we conducted a case study to better understand the performance of \logmean, and concluded that 
the \logmean model tends to have better performance if the testing search term is  frequently searched  
and has appeared in a same search session before. 

\section*{Acknowledgements}

This project was made possible, in part, by support from the National
Science Foundation under Grant Number IIS-1855501 and
IIS-1827472, and from National Library of Medicine under Grant Number 1R01LM012605-01A1. 
Any opinions, findings, and conclusions or recommendations
expressed in this material are those of the authors and do
not necessarily reflect the views of the funding agencies.



\ifCLASSOPTIONcaptionsoff
  \newpage
\fi\

\end{document}